\documentclass[twocolumn]{aastex63}

\usepackage{lineno}

\let\pwiflocal=\iffalse \let\pwifjournal=\iffalse
\usepackage[utf8]{inputenc}
\usepackage{textcomp}
\usepackage{amsmath,amssymb,gensymb,times,graphicx,morefloats,color}
\usepackage{afterpage, bm}
\usepackage{natbib,hyperref}
\usepackage[outdir=./]{epstopdf}
\usepackage{mathrsfs}
\usepackage{mathtools}
\usepackage{url}
\usepackage{rotating}
\newcommand{\teff}{\ensuremath{T_{\text{eff}}}}
\newcommand{\logg}{log$(g)$}
\newcommand{\vmac}{$v_{\rm mac}$}
\newcommand{\hal}{H$\alpha$}
\newcommand{\fbol}{$F_{\mathrm{bol}}$}
\newcommand{\rsun}{$R_\odot$}

\newcommand{\msun}{$M_\odot$}

\newcommand{\vsini}{$v$sin$i$}

\newcommand{\lii}{Li {\scshape i}}

\newcommand{\kms}{km s$^{-1}$}
\newcommand{\gaia}{{\it Gaia}}
\newcommand{\kepler}{{\it Kepler}}

\newcommand{\tess}{{\it TESS}}

\newcommand{\fspot}{$f_{\rm s}$}
\newcommand{\fspotecl}{$f_{\rm s, ecl}$}

\received{Sep 15, 2022}
\revised{Nov 22, 2022}
\accepted{Nov 23, 2022}

\shortauthors{Tofflemire et al.}
\shorttitle{TOI 450: A 40 Myr Eclipsing Binary}

\begin{document}

\title{A Low-mass, Pre-main-sequence Eclipsing Binary in the 40 Myr Columba Association -- Fundamental Stellar Parameters and Modeling the Effect of Star Spots}

\correspondingauthor{Benjamin M.\ Tofflemire}
\email{tofflemire@utexas.edu}

\author[0000-0003-2053-0749]{Benjamin M. Tofflemire}
\altaffiliation{51 Pegasi b Fellow}
\affiliation{Department of Astronomy, The University of Texas at Austin, Austin, TX 78712, USA}

\author[0000-0001-9811-568X]{Adam L.\ Kraus}
\affiliation{Department of Astronomy, The University of Texas at Austin, Austin, TX 78712, USA}

\author[0000-0003-3654-1602]{Andrew W.\ Mann}
\affiliation{Department of Physics and Astronomy, The University of North Carolina at Chapel Hill, Chapel Hill, NC 27599, USA} 

\author[0000-0003-4150-841X]{Elisabeth R.\ Newton}
\affiliation{Department of Physics and Astronomy, Dartmouth College, Hanover, NH 03755, USA}
\affiliation{Department of Physics and Kavli Institute for Astrophysics and Space Research, Massachusetts Institute of Technology, Cambridge, MA 02139, USA}

\author[0000-0002-4020-3457]{Michael A.\ Gully-Santiago}
\affiliation{Department of Astronomy, The University of Texas at Austin, Austin, TX 78712, USA}

\author[0000-0001-7246-5438]{Andrew Vanderburg}
\affiliation{Department of Physics and Kavli Institute for Astrophysics and Space Research, Massachusetts Institute of Technology, Cambridge, MA
02139, USA}

\author[0000-0002-8961-0352]{William~C.~Waalkes}
\altaffiliation{NSF Graduate Research Fellow}
\affiliation{Department of Astrophysical \& Planetary Sciences, University of Colorado Boulder, 2000 Colorado Ave, Boulder, CO 80309, USA}

\author[0000-0002-3321-4924]{Zachory K.\ Berta-Thompson}
\affiliation{Department of Astrophysical \& Planetary Sciences, University of Colorado Boulder, 2000 Colorado Ave, Boulder, CO 80309, USA}

\author[0000-0003-2781-3207]{Kevin I. Collins} 
\affiliation{George Mason University, 4400 University Drive, Fairfax, VA, 22030 USA}

\author[0000-0001-6588-9574]{Karen A.\ Collins}
\affiliation{Center for Astrophysics \textbar \ Harvard \& Smithsonian, 60 Garden Street, Cambridge, MA 02138, USA}

\author[0000-0002-5254-2499]{Louise D. Nielsen}
\affiliation{European Southern Observatory, Karl-Schwarzschild-Stra{\ss}e 2, 85748 Garching bei M{\"u}nchen, Germany} 

\author[0000-0002-7613-393X]{Fran{\c c}ois Bouchy}
\affiliation{Departement d'astronomie, Université de Gen{\'e}ve, Chemin Pegasi, 51, CH-1290 Versoix, Switzerland}

\author[0000-0002-0619-7639]{Carl Ziegler}
\affiliation{Department of Physics, Engineering and Astronomy, Stephen F. Austin State University, 1936 North St, Nacogdoches, TX 75962, USA}

\author[0000-0001-7124-4094]{C\'{e}sar Brice\~{n}o}
\affiliation{Cerro Tololo Inter-American Observatory, Casilla 603, La Serena, Chile}

\author[0000-0001-9380-6457]{Nicholas M.\ Law}
\affiliation{Department of Physics and Astronomy, The University of North Carolina at Chapel Hill, Chapel Hill, NC 27599, USA} 

\begin{abstract}

Young eclipsing binaries (EBs) are powerful probes of early stellar evolution. Current models are unable to simultaneously reproduce the measured and derived properties that are accessible for EB systems (e.g., mass, radius, temperature, luminosity). In this study we add a benchmark EB to the pre-main-sequence population with our characterization of TOI 450 (TIC 77951245). Using \gaia\ astrometry to identify its comoving, coeval companions, we confirm TOI 450 is a member of the $\sim$40 Myr Columba association. This eccentric ($e=0.2969$), equal-mass ($q=1.000$) system provides only one grazing eclipse. Despite this, our analysis achieves the precision of a double-eclipsing system by leveraging information in our high-resolution spectra to place priors on the surface-brightness and radius ratios. We also introduce a framework to include the effect of star spots on the observed eclipse depths. Multicolor eclipse light curves play a critical role in breaking degeneracies between the effects of star spots and limb-darkening. Including star spots reduces the derived radii by $\sim$2\% from an unspotted model ($>2\sigma$) and inflates the formal uncertainty in accordance with our lack of knowledge regarding the star spot orientation. We derive masses of 0.1768($\pm$0.0004) and 0.1767($\pm$0.0003) $M_\odot$, and radii of 0.345($\pm$0.006) and 0.346($\pm$0.006) $R_\odot$ for the primary and secondary, respectively. We compare these measurements to multiple stellar evolution isochones, finding good agreement with the association age. The MESA MIST and SPOTS ($f_{\rm s}=0.17$) isochrones perform the best across our comparisons, but detailed agreement depends heavily on the quantities being compared.

\

\end{abstract}

\section{Introduction} 

Research on the formation and evolution of low-mass stars and planets relies on fundamental stellar parameters derived from stellar evolution models. As with many subfields of astrophysics, theoretical stellar models provide a foundation for addressing many of our most pressing open questions. Often, the fundamental parameter in question is age, shaping our understanding pre-main-sequence (pre-MS) stellar evolution \citep{Stassunetal2014,Davidetal2019}, age-activity relations \citep{Preibischetal2005,Pace2013}, and gyrochronology \citep{Barns2007,Mamajek&Hillenbrand2008,Rebulletal2016}, while also breaking the age--mass degeneracy for directly imaged giant planets \citep[e.g.,][]{Hinkleyetal2013}. With mass, we can characterize the initial mass function \citep{Bastianetal2010}. With radius, we can derive the radii of transiting planets \citep{Gaidosetal2012}, which is particularly exciting at young ages where planets are expected to evolve through some combination of thermal contraction \citep{Fortneyetal2011}, photoevaporation \citep{Owen&Jackson2012,Owen&Wu2013}, and core-powered \citep{Ginzburgetal2018} mass loss. 

Despite their far-reaching application, there exist few direct tests of the accuracy of fundamental parameters predicted by models, especially at young ages. This has led to the development of (semi)empirical relations \citep[e.g.,][]{Torresetal2010,Mannetal2015a,Mannetal2019,Kesselietal2019} to avoid the systematic uncertainties that accompany model-dependent values. Empirical relations are widespread for main sequence (MS) stars but are sparse at young ages \citep{Herczeg&Hillenbrand2014,Davidetal2019}. Benchmarking stellar evolution models at young ages is an important step in developing accurate models, including identifying the physical processes that are missing.

Detached eclipsing binaries (EBs) are one pathway for benchmarking stellar models. The fortuitous orientation in which we view these systems allows for the measurement of their masses and radii at statistical uncertainties that routinely reach better than 1\% precision. This precision far surpasses what is possible for single stars and, critically, EB measurements rely on few model-dependent assumptions, making them less susceptible to the typical inherited systematic uncertainties. When an EB is a member of young association or cluster, additional high-precision measurements are afforded from the coeval ensemble (e.g., age, metallicity). 

EBs have a long history of testing stellar evolution theory \citep[e.g.,][and references therein]{Andersen1991}. A primary finding is that models consistently underestimate MS stellar radii by $\sim$5\% \citep{Lopez-Morales2007,Torresetal2010}. The most common hypothesis for the discrepancy is the effect of magnetic activity. Short-period EBs are expected, and observed, to have high activity levels due to rapid rotation from tidal spin-up by their binary companions \citep{Krausetal2011a}. However, a similar level of discrepancy exists for long-period systems \citep{Irwinetal2011}. Magnetic fields have been implemented in stellar models in their ability to inhibit convective flows \citep{Feiden&Chaboyer2012,Feiden&Chaboyer2013}, and to alter standard radiative transfer via star spots \citep{Somers&Pinsonneault2015,Somersetal2020}. 

While the inclusion of magnetic field prescriptions appears to ease the tension for MS stars, discrepancies exist on larger scales for pre-MS stars, particularly at low masses. In the study of nine EBs in the 5--7 Myr Upper Sco association, \citet{Davidetal2019} found there is good relative agreement among most models between 0.3 and 1 $M_\odot$, but that they {\it overpredict} the radii for young stars below 0.3 $M_\odot$. This is the opposite of the MS radius discrepancy, highlighting that, although magnetic fields are likely altering these young systems in similar ways to MS stars, larger-scale uncertainties exist in our understanding of pre-MS evolution.

Beyond the shortcomings of current models, which are likely due, in part, to the absence of magnetic phenomena, the observational characterization of EBs typically also ignores their effects. EB analyses rely on few model assumptions, but one common assumption is that stellar photospheres can be described as a uniform, limb-darkened disk. This assumption is false for any young system where star spots are not only present, but likely have large covering fractions \citep{Gully-Santiagoetal2017,Fangetal2018,Caoetal2022}. The specific orientation of spots or spot complexes alters the detailed surface-brightness distributions, and can significantly impact the measured eclipse depths \citep{Moralesetal2010,Rackhametal2018}. The direction and magnitude of this effect depend on the specific spot geometries with respect to the eclipse geometry, and are unlikely to result in a consistent systematic offset common to all EB radius measurements. Still, given that spot geometries are rarely known and their effects are rarely addressed in eclipse light-curve modeling, quoted radii uncertainties (often $\lesssim$1\%) are likely underestimated for spotted systems. This underestimation of the error may be a contributing factor to the significant discrepancies found in the derived radii between different groups modeling the same EB systems (e.g., see \citealt{Moralesetal2009} and \citealt{Windmilleretal2010}; \citealt{Krausetal2017} and \citealt{Gillenetal2017}). 

\begin{figure*}[th!]
\begin{center}
\includegraphics[width=0.98\textwidth]{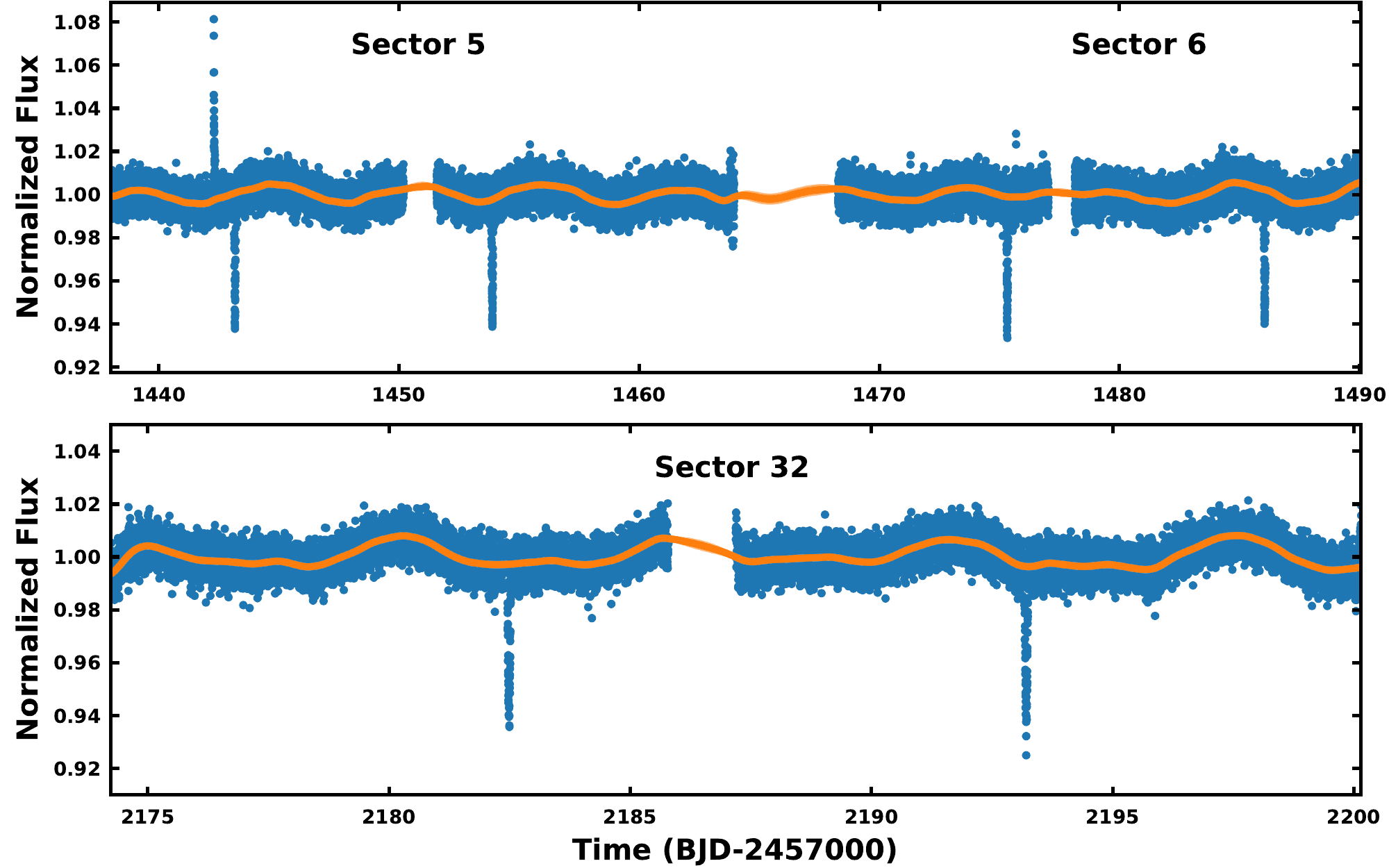}
\caption{\tess\ light curve of TOI 450. The top panel displays Sectors 5 and 6, with data from Sector 32 in the bottom panel. The Gaussian process variability model is shown in the orange.}
\label{fig:tess}
\end{center}
\end{figure*}

As part of an effort to increase the population of young, benchmark EBs, we present the characterization of TOI 450 (TIC 77951245). Initial followup of the nominal planet host was undertaken by the THYME collaboration \citep[\tess\ Hunt For Young and Maturing Exoplanets;][]{Newtonetal2019} and the \tess\ Follow-up Observing Program (TFOP) community, where it was identified as a double-lined spectroscopic binary (SB2) \citep{Battleyetal2020}. In this study, we confirm TOI 450's membership to the $\sim$40 Myr Columba association using the kinematic selection methodology presented in \citet{Tofflemireetal2021}, now updated for \gaia\ DR3 \citep{GAIA2016,Gaiadr3}. 

We then perform a joint radial-velocity (RV) and eclipse light-curve fit to derive the fundamental parameters of the system, confirming its components are on the pre-MS. Our analysis includes two key additions to standard EB modeling. First we place a joint prior on the surface-brightness ratio and radius-ratio informed by our spectroscopic decomposition. This prior enables a fit to this single-eclipsing system that reaches a formal precision on par with double-eclipsing systems. Second, we develop and implement a framework to include the effect star spots have on eclipse depths. Our ability to constrain the impact of spots relies heavily on multicolor eclipse observations. The combination of \tess\ to find EBs and \gaia\ to confirm their association memberships, and therefore age, makes this a pivotal time in our ability to find benchmark EBs and improve our understanding of early stellar evolution.

\section{Observations \& Data Reduction}

\subsection{Time-Series Photometry}

\subsubsection{TESS}

TOI 450 was observed by \tess\ \citep{Rickeretal2015} with 2 min cadence during Sectors 5 and 6 in Cycle 1 of the primary mission (UT Nov 15, 2018 -- Jan 6, 2019), and during Sector 32 of the extended mission (UT Nov 20, 2020 --  Dec 16, 2020). In all observations, TOI 450 fell on Camera 3. Two-minute cadence data are processed by the SPOC pipeline \citep{Jenkins2015,Jenkinsetal2016}. Our analysis makes use of the presearch data conditioning simple aperture photometry \citep[PDCSAP;][]{SmithKepler2012,StumpeKepler2012,StumpeMultiscale2014} light curve. 

Figure \ref{fig:tess} presents the \tess\ light curves, where two clear eclipse events can be seen in each sector. The light curve also shows stellar flares, seen most clearly at the beginning of Sector 5, and spot modulation. The eclipse events were detected by the SPOC Transiting Planet Search pipeline \citep[TPS;][]{Jenkins2002,Jenkinsetal2010} with a period of 10.71 days and was alerted as a \tess\ Object of Interest (TOI), TOI 450, in May 2019 \citep{Guerreroetal2021}.

\subsubsection{Las Cumbres Observatory Global Telescope -- 1.0 m Network}

Follow-up eclipse monitoring was performed with the Las Cumbres Observatory Global Telescop (LCOGT) 1.0\,m telescope network \citep{Brownetal2013}. All thirteen 1-m telescopes are outfitted with 4096$\times$4096 pixel Sinistro CCD imagers ($0\farcs39$ pixel$^{-1}$). Raw images are reduced with the LCO BANZAI pipeline \citep{McCullyetal2018} and photometric data are extracted with {\tt AstroImageJ} \citep{Collinsetal2017}. 

One full eclipse was successfully monitored on 2019 Feb 25 UTC. These observations were completed with two 1 m telescopes at the Cerro Tololo Inter-American Observatory (CTIO) in the Sloan $r'$ and $I$ filters. The observations were 224 and 208 minutes in duration, centering on the eclipse, with effective cadences of 188 and 60 seconds, respectively. Differential photometry was computed using eight and five nonvarying field stars, respectively. The final differential light curve includes airmass detrendeing. 

\

\subsection{Spectroscopy}

\subsubsection{SALT--HRS}

During the fall of 2019, 11 epochs of high-resolution optical spectra were obtained with the High Resolution Spectrograph \citep[HRS;][]{Crauseetal2014} on the Southern African Large Telescope \cite[SALT;][]{Buckleyetal2006} located at the South African Astronomical Observatory. HRS is a cross-dispersed echelle spectrograph with separate blue and red arms that cover a 3700--8900 \AA. Our observations were made in the high-resolution mode, which delivers an effective resolution of $R\sim46,000$. Data reduction, flat field correction, and wavelength calibration are performed with the facility's MIDAS pipeline \citep{Kniazevetal2016,Kniazevetal2017}. For each epoch, three spectra were taken back-to-back and reduced individually. Table \ref{tab:rv} presents the mean BJD of each epoch and our RV measurements (see Section \ref{rvs}). 

\subsubsection{ESO 3.6m--HARPS}

TOI 450 was observed three times in the fall of 2019 with the HARPS spectrograph \citep{Mayoretal2003} on the ESO 3.6m telescope in the high-efficiency mode (EGGS) as part of the follow-up efforts of NGTS planet candidates \citep[NOI-104351;][]{Wheatleyetal2018}. 
These spectra cover a wavelength range of 3782–6913 \AA\ at a spectral resolution of R$\sim$80,000. Monitoring was stopped after the target was identified as an SB2. We derive RV measurements from them, and provide their relevant information in Table 2.

\subsection{Speckle Imaging: SOAR--HRCam}
\label{speckle}

\begin{figure}[h!]
\begin{center}
\includegraphics[width=0.47\textwidth]{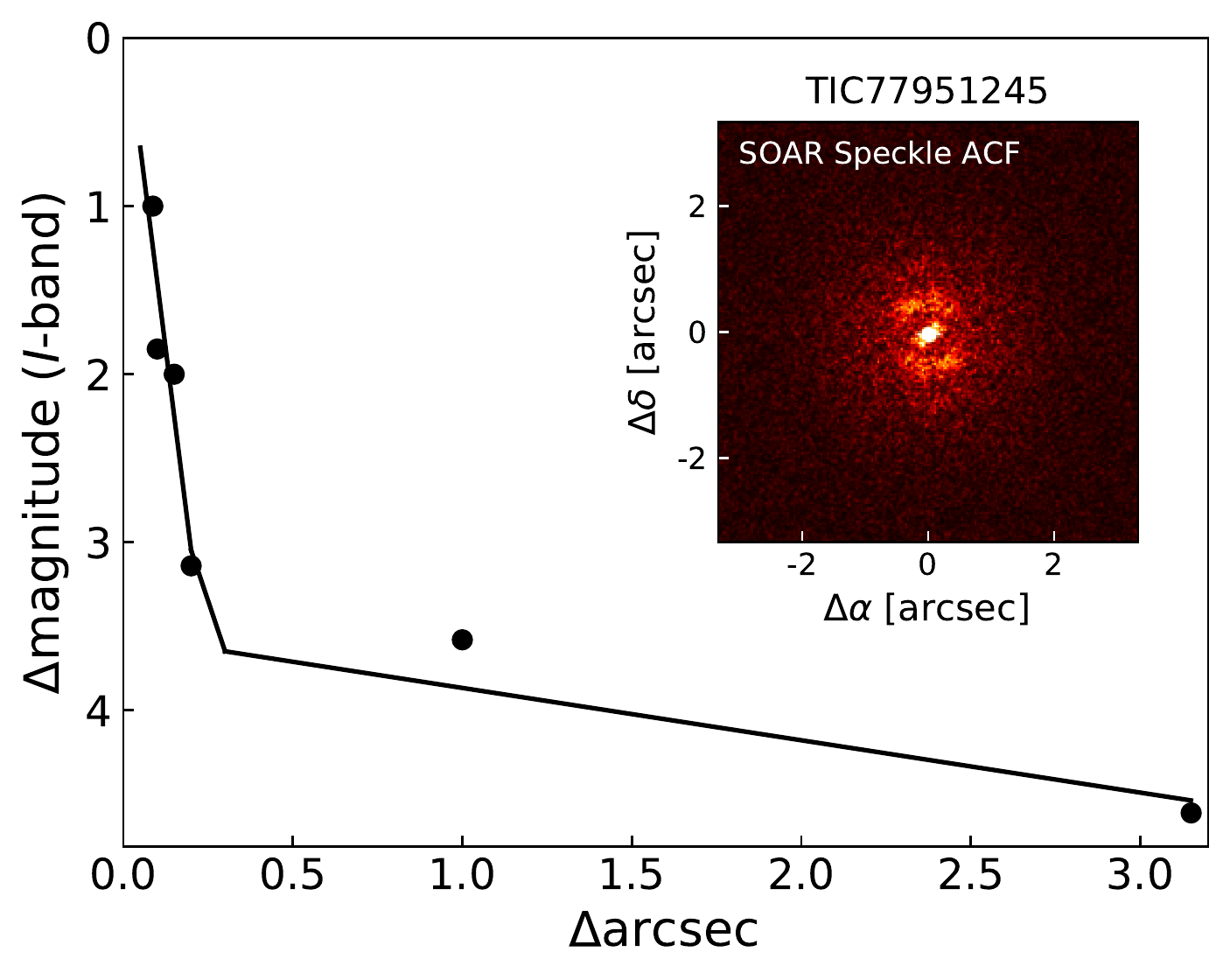}
\caption{Detection limits (5$\sigma$) for companions to TOI 450 as observed by the SOAR HRCam speckle imager. The reconstructed image is presented in the inset panel. No companions are detected. }
\label{fig:HRCam}
\end{center}
\end{figure}

Speckle imaging of TOI 450 was obtained to assess the presence of unresolved companions, which can alter the color and depth of eclipses. Our observations were made on 2019 Mar 17 (UTC) with the High-Resolution Camera (HRCam) on the 4.1 m Southern Astrophysical Research (SOAR) telescope. Observations were made in the $I$-band ($\lambda_{\rm eff}\sim$ 8790 \AA). Details on HRCam observations and data reduction, as well as the SOAR \tess\ survey are described in  \citet{Ziegleretal2020}. 
Figure \ref{fig:HRCam} presents the 5$\sigma$ contrast curve, where no sources are detected within 3$\arcsec$. Adopting the $\tau = 40$ Myr isochrones of \citet{Baraffeetal2015}, the corresponding limits in companion mass and physical projected separation are $M < 85 M_{Jup}$ at $\rho = 5.3$ AU, $M < 55 M_{Jup}$ at $\rho = 8.0$ AU, $M < 40 M_{Jup}$ at $\rho = 10.6$ AU, and $M < 35 M_{Jup}$ at $\rho \ge 16$ AU.

\subsection{Limits on Companions from Gaia EDR3}

The presence of nearby companions can inflate the astrometric errors in \gaia\ observations, resulting a larger value of the Renormalized Unit Weight Error \citep[RUWE;][]{Lindegrenetal2018} above the expected value of $RUWE=1.0$ for a star with a well-behaved astrometric solution. This inflation can result from genuine photocenter orbital motion that is not yet being modeled \citep[e.g.,][]{Belokurovetal2020} or from the influence of spatially resolved companions that bias the centroid measurements (\citealt{Rizzutoetal2018}; \citealt{Woodetal2021}; A.\ L.\ Kraus et al., in prep). The Gaia documentation recommends a threshold of $RUWE = 1.4$ for assessing whether the astrometry is being inflated, but the RUWE distribution of old field stars suggests that $RUWE = 1.2$ provides a robust discriminator for field stars (\citealt{Brysonetal2020a}; A.\ L.\ Kraus et al., in prep). However, the distribution is biased to higher values of RUWE for known single stars in young stellar populations \citep[$\sim$10; Myr][]{Fittonetal2022}. RUWE might be inflated in protoplanetary disk hosts due to scattered light (with a 95\% threshold of $RUWE = 2.5$), but also in young disk-free stars, perhaps due to second-order effects in astrometric correction terms due to brightness or color variations (with a 95\% threshold of $RUWE = 1.6$).

In Gaia EDR3, TOI 450 seems to have mildly inflated astrometric scatter ($RUWE = 1.324$) with respect to the estimated uncertainties. This value would represent an excess with respect to well-behaved field stars, but does not exceed the threshold generically suggested for all sources by the Gaia team, nor the threshold seen for young disk-free stars by \citet{Fittonetal2022}. There is no evidence of additional companions from speckle imaging (Section \ref{speckle}) or followup spectroscopy (Section \ref{spec_comps}), so the mild RUWE excess should not be regarded as strong evidence of any additional companions in the system. 

Finally, the Gaia EDR3 catalog also provides deep limits on additional companions within the system. The membership of this system in Columba implies that there will be very wide comoving neighbors, but there are no comoving and codistant sources in the Gaia EDR3 catalog within $\rho < 1900 \arcsec$ ($\rho \la 10^5$ AU). Nearby sources typically have five-parameter solutions if brighter than $G < 20.7$ mag ($M \ga 15$ $M_{Jup}$ at $\tau \sim 40$ Myr; \citealt{Baraffeetal2015}). We therefore conclude that there are no wide stellar or brown dwarf companions to TOI 450.

\subsection{Literature Photometry \& Astrometry}

We compile broadband photometry and astrometry from various surveys in our characterization of the TOI 450 system (Sections \ref{sed} and \ref{priors}) and our assessment of its membership to the Columba moving group (Section \ref{ff-columba}). Table \ref{tab:stellar_props} compiles these measurements and other relevant quantities we derive from them. 

\begin{deluxetable}{l c c}
\tablecaption{Properties of TOI 450
\label{tab:stellar_props}}
\tablewidth{0pt}
\tabletypesize{\footnotesize}
\tablecolumns{3}
\phd
\tablehead{
  \colhead{Parameter} &
  \colhead{Value} &
  \colhead{Source}
}
\startdata
\multicolumn{3}{l}{\textbf{Identifiers}} \\
TOI       & 450                  & \citet{Guerreroetal2021} \\
TIC       & 77951245             & \citet{Stassunetal2018} \\
2MASS     & J05160118-3124457    & 2MASS \\
\gaia\ DR2& 4827527233363019776  & \gaia\ DR2 \\
\gaia\ EDR3& 4827527233363019776 & \gaia\ EDR3 \\
\hline
\multicolumn{3}{l}{\textbf{Astrometry}} \\
$\alpha$ RA (J2000)          & 05:16:01.179534 & \gaia\ EDR3\\
$\delta$ Dec (J2000)         & $-$31:24:45.6858 & \gaia\ EDR3\\
$\mu_\alpha$ (mas yr$^{-1}$) & 34.286 $\pm$ 0.018 & \gaia\ EDR3\\
$\mu_\delta$ (mas yr$^{-1}$) & $-$0.794 $\pm$ 0.019 & \gaia\ EDR3\\
$\pi$ (mas)                  & 18.649 $\pm$ 0.018 & \gaia\ EDR3\\
$RUWE$                       & 1.324 & \gaia\ EDR3\\
\hline
\multicolumn{3}{l}{\textbf{Photometry}} \\
{\it B} (mag) & 16.7 $\pm$ 0.4 & APASS DR9 \\
{\it V} (mag) & 15.2 $\pm$ 0.2 & APASS DR9 \\
{\it G$_{\rm BP}$} (mag)  & 15.560 $\pm$ 0.005 & \gaia\ EDR3\\
{\it G} (mag)   & 13.782 $\pm$  0.003 & \gaia\ EDR3\\
{\it G$_{\rm RP}$} (mag)  & 12.511 $\pm$ 0.004 & \gaia\ EDR3\\
{\it J} (mag)   & 10.63 $\pm$ 0.03 & 2MASS\\
{\it H} (mag)   & 10.14 $\pm$ 0.02 & 2MASS\\
{\it K$_s$} (mag)   & 9.79 $\pm$ 0.02 & 2MASS\\
{\it W}1 (mag)  & 9.60 $\pm$ 0.02 & {\it WISE}\\
{\it W}2 (mag)  & 9.43 $\pm$ 0.02 & {\it WISE}\\
{\it W}3 (mag)  & 9.27 $\pm$ 0.03 & {\it WISE}\\
{\it W}4 (mag)  & 8.92 $\pm$ 0.42 & {\it WISE}\\
\hline
\multicolumn{3}{l}{\textbf{Kinematics \& Positions}} \\
RV (\kms)       & 23.7 $\pm$ 0.5 & This Work\\
U (\kms)        & -12.40 $\pm$ 0.03   & This Work\\
V (\kms)        & -21.23 $\pm$ 0.04   & This Work\\
W (\kms)        & -5.90 $\pm$ 0.03   & This Work\\
X (pc)          & -26.12 $\pm$ 0.02   & This Work \\
Y (pc)          & -36.45 $\pm$ 0.03   & This Work \\
Z (pc)          & -29.14 $\pm$ 0.03   & This Work \\
Distance (pc)   & 53.48 $\pm$ 0.05 & \citet{Bailer-Jonesetal2021} \\
\enddata
\end{deluxetable}

\section{Analysis}

In this section we describe the analysis of our primary data sets. These measurements serve as inputs to our joint RV and eclipse fit in Section \ref{ebfit} and provide important priors that enable a precise analysis of this grazing EB system. 

\subsection{Radial Velocities}
\label{rvs}

\begin{figure}[t!]
\begin{center}
\includegraphics[width=0.47\textwidth]{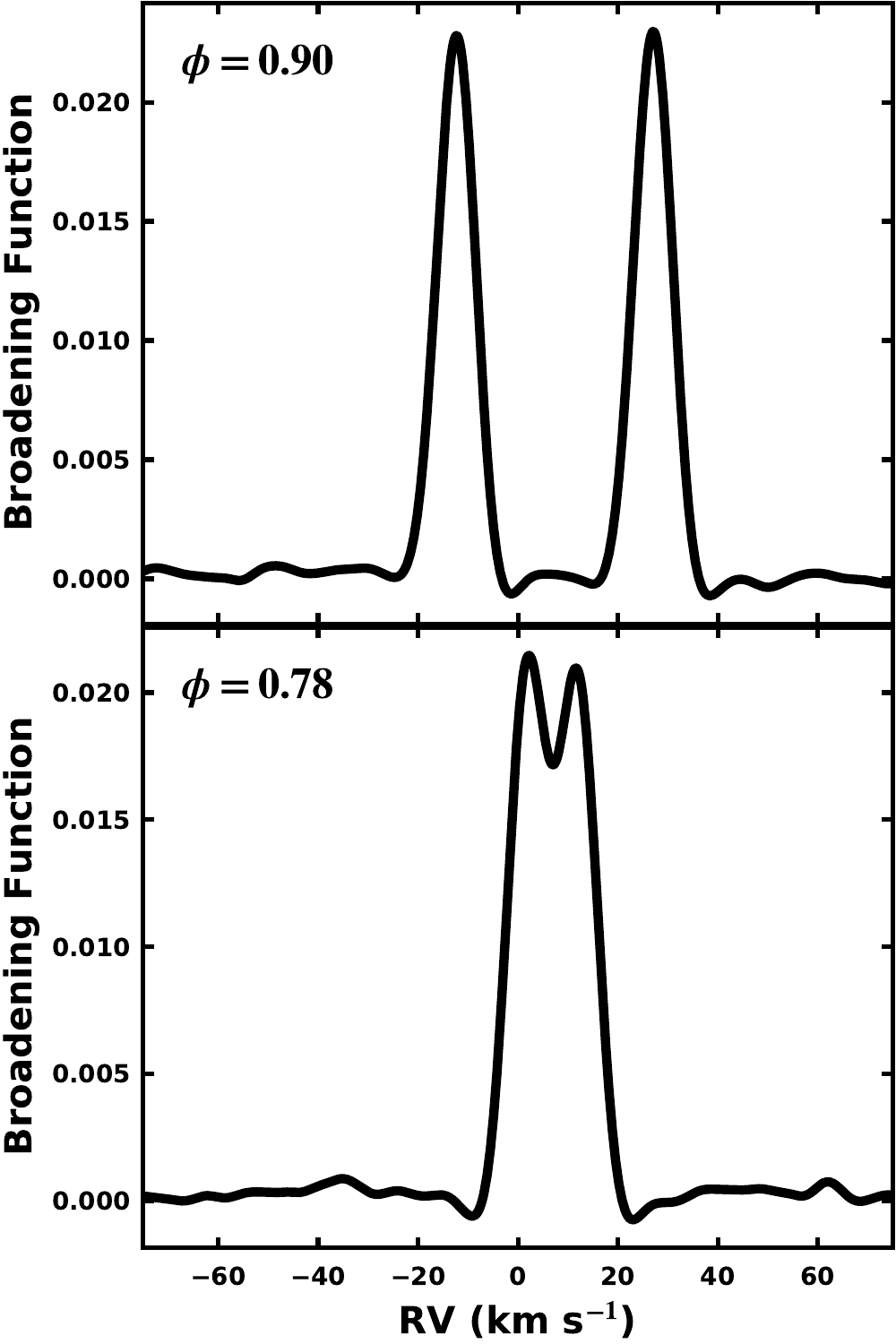}
\caption{Combined Broadening Function for two epochs of SALT--HRS spectra. These highlight a typical epoch that is well separated (top) and blended (bottom). The orbital phase of each epoch is provided in the top-left of the panel. }
\label{fig:bf}
\end{center}
\end{figure}

Stellar RVs are measured from our high-resolution optical spectra by computing spectral line broadening functions (BFs; \citealt{Rucinski1992}) using the {\tt saphires} python package \citep{Tofflemireetal2019}. The BF is the result of a linear inversion of an observed spectrum with a narrow-lined template, and represents a reconstruction of the average stellar absorption-line profile. When the observed spectrum contains the light from two stars, as it does in an SB2 system like TOI 450, the BF provides the velocity profile of each star. Figure \ref{fig:bf} displays the BFs from two epochs. The BF is similar to the commonly used cross-correlation function (CCF), but offers a higher fidelity result \citep{Rucinski1999b}\footnote{Example of a CCF and BF comparison: \url{http://www.astro.utoronto.ca/~rucinski/SVDcookbook.html}} whose profiles more directly map to physical properties (e.g., \vsini, flux ratio). The higher fidelity, in particular, is critical when decomposing blended stellar profiles common in SB2 observations. 

Synthetic spectra generally make poor narrow-lined templates, especially in the case of low-mass stars where the detailed match with observations at high resolution is still limited. Empirical templates produce BFs with much lower noise due to their improved match. The trade-off is that empirical-template BFs no longer reproduce the average absorption-line profile, but rather the profile that will reproduce the observed spectrum when convolved with the template. The result is a narrower BF profile, which aids in RV precision. As this is the goal of the current analysis, we create empirical spectral templates for spectral types M0.0 through M5.0 in steps of 0.5 using the CARMENES spectral library \citep{Reinersetal2018}. Only slowly rotating (unresolved line profiles; \vsini\ $<$ 2 \kms) stars are included. Using a uniform cubic basic (B-spline) regression following the {\tt SERVAL} package's implementation \citep{Zechmeisteretal2018}, we a create spectral template for each order, oversampling the spline to match the native number of resolutions elements in the order.  We find consistent results (RVs) across the spectral templates, but find the M4.5 template produces the consistently highest signal-to-noise BFs from order to order. As such, we adopt it as our narrow-lined template. 

With our M4.5 narrow-lined template, we compute the BF for individual SALT--HRS orders with high signal to noise and low telluric contamination. In practice, this includes 34 orders from $\sim5200-8800$ \AA. For the HARPS spectra, we break the 1D spectrum (default data product) into 26 sections of $\sim$ 100 \AA\ in length, covering $\sim5200-6700$ \AA. Individual orders are then combined into a high signal to noise BF, weighted by the noise at high velocities where no stellar contributions are present. For 10 of our 14 spectra, the stellar components do not overlap in velocity space (e.g., botton panel of Figure \ref{fig:bf}). Each component is fit with a Gaussian profile to measure the stellar RV. Uncertainty on the RV measurement is assessed with a bootstrap approach in which $10^5$ BFs are combined and fit from a random sampling with replacement of the contributing orders. The standard deviation of the RV measurement distribution is adopted as the uncertainty. For four epochs where the stellar profiles are blended (e.g., Figure \ref{fig:bf} bottom), we impose bounds on the relative strength of the two fit components, informed by the $3\sigma$ bounds of the values measured in well-separated epochs. This bound prevents nonphysical flux-ratio values (see Section \ref{FR}) that can skew the RV values. For the SALT--HRS epochs, we adopt a weighted mean and standard deviation of the three individual spectra as our value. Observed RVs are corrected to the barycentric frame using the {\tt barycorrpy} package \citep{KanodiaAndWright2018}. Our barycentric RVs and their relative uncertainty are presented in Table \ref{tab:rv}. The absolution precision of the RV measurements is on the order of 0.5 \kms, based on the offset we measure between the SALT--HRS and HARPS velocity zero-points (Section \ref{results}). 

\begin{deluxetable}{l c c c c}
\tablecaption{Radial Velocities from High-Resolution Optical Spectra
\label{tab:rv}}
\tablewidth{0pt}
\tabletypesize{\footnotesize}
\tablecolumns{5}
\phd
\tablehead{
  \colhead{Facility} &
  \colhead{BJD} &
  \colhead{$RV_1$} &
  \colhead{$RV_2$} &
  \colhead{Orbital Phase$^a$} \\
  \colhead{} &
  \colhead{} &
  \colhead{(\kms)} &
  \colhead{(\kms)} &
  \colhead{}
}
\startdata
HARPS & 2458693.91849 & 25.85  $\pm$ 0.16 & 23.01 $\pm$ 0.15 & 0.31 \\
SALT  & 2458706.65220 & 45.44  $\pm$ 0.05 & 2.57  $\pm$ 0.12 & 0.50 \\
SALT  & 2458708.64979 & 50.20  $\pm$ 0.06 & -2.19 $\pm$ 0.11 & 0.69 \\
SALT  & 2458721.61189 & 19.26  $\pm$ 0.28 & 28.67 $\pm$ 0.28 & 0.90 \\
SALT  & 2458734.57612 & -14.55 $\pm$ 0.18 & 62.20 $\pm$ 0.20 & 0.11 \\
SALT  & 2458744.55144 & -21.02 $\pm$ 0.20 & 67.94 $\pm$ 0.06 & 0.04 \\
SALT  & 2458752.52819 & 43.74  $\pm$ 0.13 & 4.11  $\pm$ 0.05 & 0.78 \\
SALT  & 2458754.52701 & -6.00  $\pm$ 0.23 & 54.20 $\pm$ 0.37 & 0.97 \\
SALT  & 2458760.50590 & 47.20  $\pm$ 0.10 & 1.11  $\pm$ 0.12 & 0.53 \\
SALT  & 2458764.49671 & 18.86  $\pm$ 0.31 & 29.88 $\pm$ 0.32 & 0.90 \\
SALT  & 2458767.48917 & 0.89   $\pm$ 0.13 & 47.26 $\pm$ 0.17 & 0.18 \\
SALT  & 2458768.49413 & 19.39  $\pm$ 0.16 & 28.82 $\pm$ 0.16 & 0.27 \\
HARPS & 2458808.76673 & -19.79 $\pm$ 0.02 & 68.46 $\pm$ 0.03 & 0.03 \\
HARPS & 2458813.75456 & 45.44  $\pm$ 0.04 & 3.2   $\pm$ 0.04 & 0.49 \\
\enddata
\tablenotetext{a}{Orbital phase $\phi = 0$ corresponds to periastron passage.}
\end{deluxetable}

\subsection{Spectroscopic Components}
\label{spec_comps}

We clearly detect two stellar components in the combined BF (Figure \ref{fig:bf}), as expected for a high-mass-ratio EB. The absence of other features in the BF provides an independent limit on the presence of additional companions, bound or otherwise. Computing a quantitative limit on the detection threshold of an additional companion is not straightforward given that our sensitivity to companions depends on their spectral features (i.e., spectral type or \teff) and rotational velocity. Still, we easily detect the binary components using empirical templates $\sim$4 spectral subtypes away from the optimal value, and similarly, \citet{Tofflemireetal2019} showed sensitivity to component detection with synthetic template mismatch of $500$ K. Furthermore, a luminous component in the spectrum with different spectral features (i.e., a much earlier spectral type) would introduce structure and noise in the high-velocity BF baseline, which is not present in our BFs for TOI 450. With this information, we can conservatively rule out the presence of slowly rotating companions (\vsini\ $<$ 10 \kms) with M spectral types and flux ratios of 10\% (2.5 mags), which would be visually obvious in the BF, within the 2\farcs2 SALT-HRS fiber. 

\begin{figure}[t!]
\begin{center}
\includegraphics[width=0.47\textwidth]{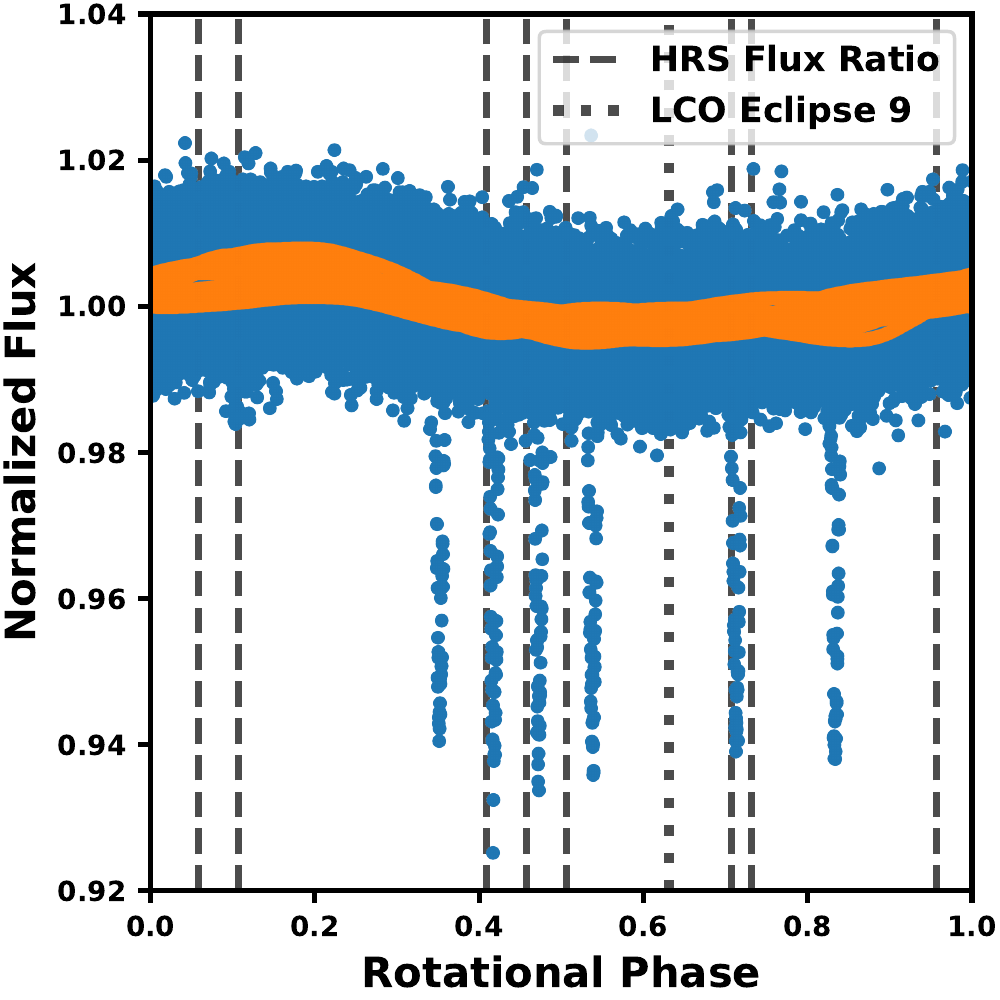}
\caption{The location of wavelength-dependent flux-ratio measurements with respect to the stellar rotational phase ($P_{rot} = 5.7$ d). Dashed lines mark the rotational phase of the eight SALT--HRS spectra with well-separated velocity components. The rotational-phase-folded \tess\ light curve from Sectors 5, 6, and 32 are shown in blue. The spot-modulation model is over-plotted in orange. Eclipse events in the \tess\ light curves are visible by eye. The dotted line shows the rotational phase of the eclipse observed with LCO, eclipse 9.}
\label{fig:FR_epochs}
\end{center}
\end{figure}

\subsection{Rotation Periods}
\label{rotp}

The \tess\ light curve contains sinusoidal modulation that results from variations in the combined, projected spot-covering fraction as each star rotates. We compute a Lomb-Scargle periodogram \citep{Scargle1982} for each \tess\ Sector (masking out the eclipse events) finding only one strong, consistent peak near 5.7 days. Smaller, yet technically significant, peaks in the periodogram likely arise from spectral leakage due to the modulation not being strictly sinusoidal. These features vary in location and strength from sector to sector and are not present in an autocorrelation function. From this analysis, we determine that only one astrophysical period can be extracted from the \tess\ light curves, which we interpret as both stars having the same rotation period. This result is expected given the equivalent stellar radii (Section \ref{ebfit}) and \vsini\ values between each component. 

To measure the rotation period in the presence of evolving spot configurations, we model the light curve with the {\tt celerite} Gaussian process \citep{Foreman-Mackeyetal2017}. The covariance kernel consists of a damped, driven, simple harmonic oscillator at the stellar rotation period and another at half the rotation period. In addition to the period, the kernel is described by the primary amplitude, $A$, the damping timescale (or quality factor) of the primary period, $Q_1$, the ratio of the primary to secondary amplitude ($A_2/A_1$), $Mix$, and the damping timescale of the secondary period ($P/2$), $Q_2$. After masking 2 hr windows centered on each eclipse and removing flares, we fit the parameters above in natural logarithmic space using {\tt emcee} \citep{Foreman-Mackeyetal2013}. Our fit employs 50 walkers. Fit convergence is established once the chain autocorrelation timescale ($\tau$) reaches a fractional change less than 5\% and the chain length exceeded 100$\tau$. Our posteriors discard the first five autocorrelation times as burn in.

Fits are made to each \tess\ Sector returning periods of $5.8\pm0.2$ d, $5.7\pm0.3$ d, and $5.6\pm0.2$ d for Sectors 5, 6 and 32, respectively. We adopt the error weighted mean and standard deviation, $5.7\pm0.1$ d, as the rotation period for each star. (We repeated this analysis with the SAP light curve reduction, as opposed to the PDCSAP reduction used elsewhere, finding consistent results with larger uncertainties.) Figure \ref{fig:FR_epochs} presents the rotational-phase-folded light curve from all three \tess\ Sectors with the variability model over-plotted. Very little evolution in the spot modulation is observed between \tess\ Sectors 5 and 32. 

We note that the synchronized stellar rotation period is shorter (i.e., more rapidly rotating) than the \citet{Hut1981} pseudo-synchronization prediction for TOI 450's orbital eccentricity ($P_{ps}\sim7$ days). Sub- and super-pseudo synchronous binaries have been observed in other young clusters \citep[e.g.,][]{Meibometal2006}, making our finding unsurprising. As a young association member with a benchmark age, TOI 450 may be a useful probe of tidal evolution theory.

\subsection{Projected Rotational Velocities}

To measure the projected rotational velocity (\vsini) of each component, we compute a separate set of BFs using a 3100 K, \logg\ = 4.5 synthetic template from the \citet{Husseretal2013} PHOENIX model suite. Although this template is a worse match to the observed spectra, its absorption lines have no rotational or instrumental broadening and therefore produce a BF whose width reflects the broadening components intrinsic to the observed stars. We fit the combined BF (following Section \ref{rvs}) with an absorption-line profile \citep{gray_book} that includes instrumental, rotational, and macroturbulent broadening (the synthetic template includes microturbulent velocity broadening). From the eight SALT--HRS epochs with large component velocity separations, we fit the \vsini\ and \vmac\ for each component, finding average values and standard deviations of: \vsini$_1=3.2\pm0.3$ \kms, \vmac$_1=2.0\pm0.3$ \kms, \vsini$_2=3.2\pm0.5$ \kms, and \vmac$_2=2.1\pm0.4$ \kms.

\subsection{Stellar Rotation Inclination}

With measurements of the \vsini, rotation period, and stellar radius (Section \ref{ebfit}), we can infer the inclination of the stellar rotation. The inclination probability distribution functions, computed following \citet{Masuda&Winn2020}, peak at $90^\circ$, but are broad with 95\% confidence intervals at 59$^\circ$ and 48$^\circ$, for the primary and secondary, respectively. This result is consistent with alignment between the stellar and orbital angular momentum vectors. 

\subsection{Spectroscopic Flux Ratio}
\label{FR}

\begin{figure}[t!]
\begin{center}
\includegraphics[width=0.47\textwidth]{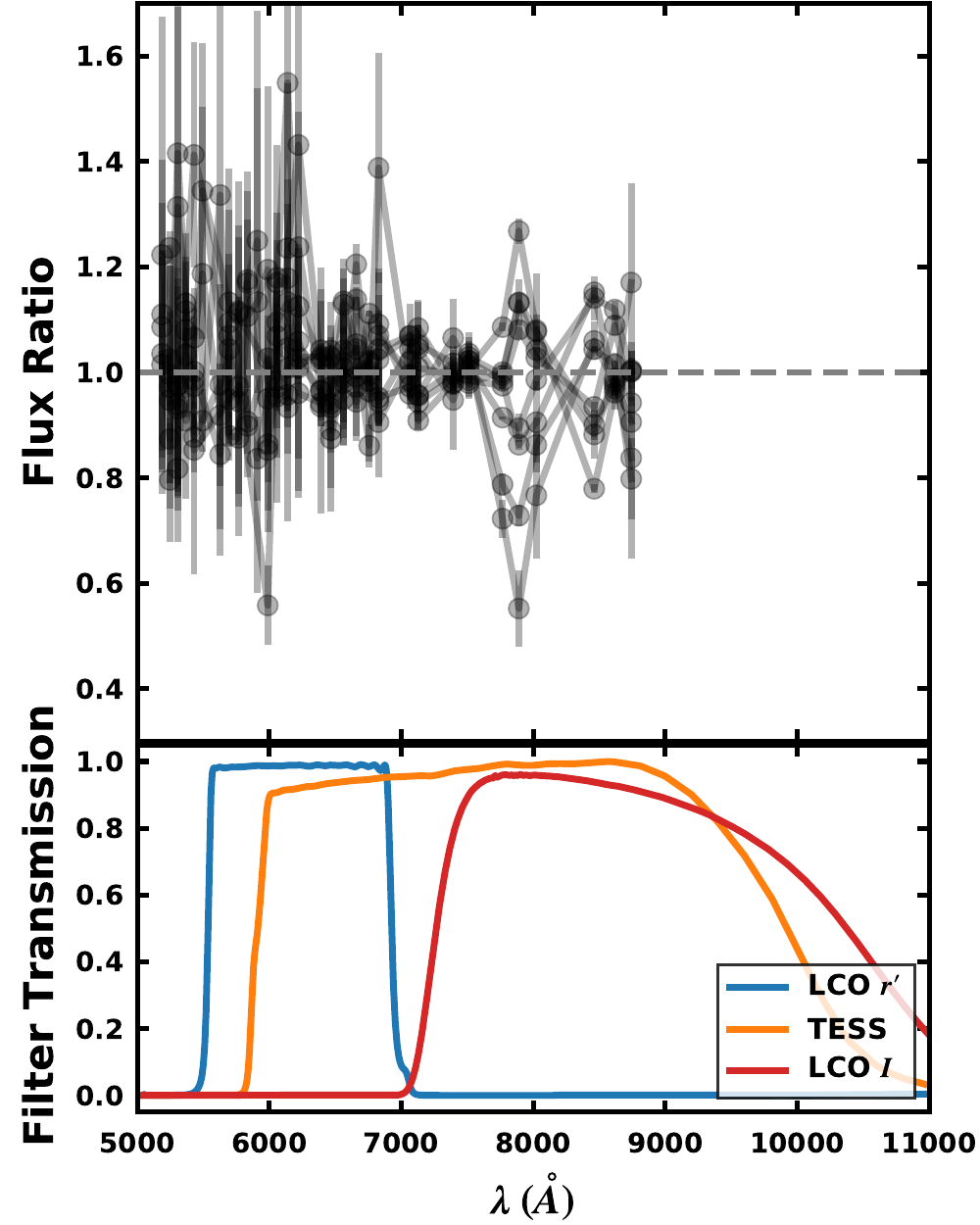}
\caption{Wavelength-dependent flux ratios from eight SALT--HRS epochs. Decreased measurement precision at short wavelengths is due to decreasing signal-to-noise at short wavelengths. Scatter in orders near $\sim$8000\AA\ probes temperature-sensitive TiO features where spot variability has the largest impact. Filter curves from the photometric filters used to observe eclipses are provided in the bottom panel. 
}
\label{fig:FR}
\end{center}
\end{figure}

In SB2 systems, the ratio of the area of the BF components encodes the flux ratio of the two stars over the wavelength range considered. For the eight SALT--HRS epochs with well-separated BF components, we measure the flux ratio for 28 orders between $\sim5200$ and 8700 \AA. Each epoch consists of three spectra, which are analyzed independently and then combined to compute the mean flux ratio and standard deviation for each order. For an order to be included for a given epoch, we demand that each of the three spectra produces a BF peak that is 5$\sigma$ above the baseline noise. This constraint removes low signal-to-noise ratio (S/N) epochs and orders.

In Figure \ref{fig:FR} we over-plot the wavelength-dependent spectroscopic flux ratios for each epoch. There is a maximum of eight epochs plotted for each order, which are presented at the order's central wavelength. Lines connect a given epoch. The $r^\prime$, \tess, and $I$ filter curves are also included for comparison. All values hover around unity with increasing uncertainty at short wavelengths as S/N decreases. The increased scatter from $\sim$7500-8500\AA\ marks orders containing temperature-sensitive TiO absorption features, which are likely influenced by the relative presence of cool spots and their variability as the stars rotate \citep{Gully-Santiagoetal2017}. 

These data capture a representative sampling of the projected surface-brightness variability over the time-baseline observed. Figure \ref{fig:FR_epochs} presents the location of our flux-ratio measurements vertical dashed lines) as a function of the stellar rotational phase (see Section \ref{rotp}). The \tess\ light curve (blue) and stellar variability model (orange) are included to provide context for the range of flux-ratio values, caused by variable projected spot-covering fractions, that our measurements probe. The spectroscopy epochs are not contemporaneous, but fall between \tess\ Sectors 6 and 32.

For TOI 450, where the system orientation only provides a single, grazing eclipse, these measurements allow for critical priors to be placed on the stellar radii and surface-brightness ratios (see Section \ref{priors}). The average flux-ratio value across all orders and epochs is $F_2/F_1 = 1.0$ with a standard deviation of 0.1. Our choice of the primary star in this system is somewhat arbitrary, but is ultimately chosen as the more massive component in our definitive fit, although both masses are the same within our uncertainty. 

\subsection{Spectral Features}

\begin{figure}[t!]
\begin{center}
\includegraphics[width=0.47\textwidth]{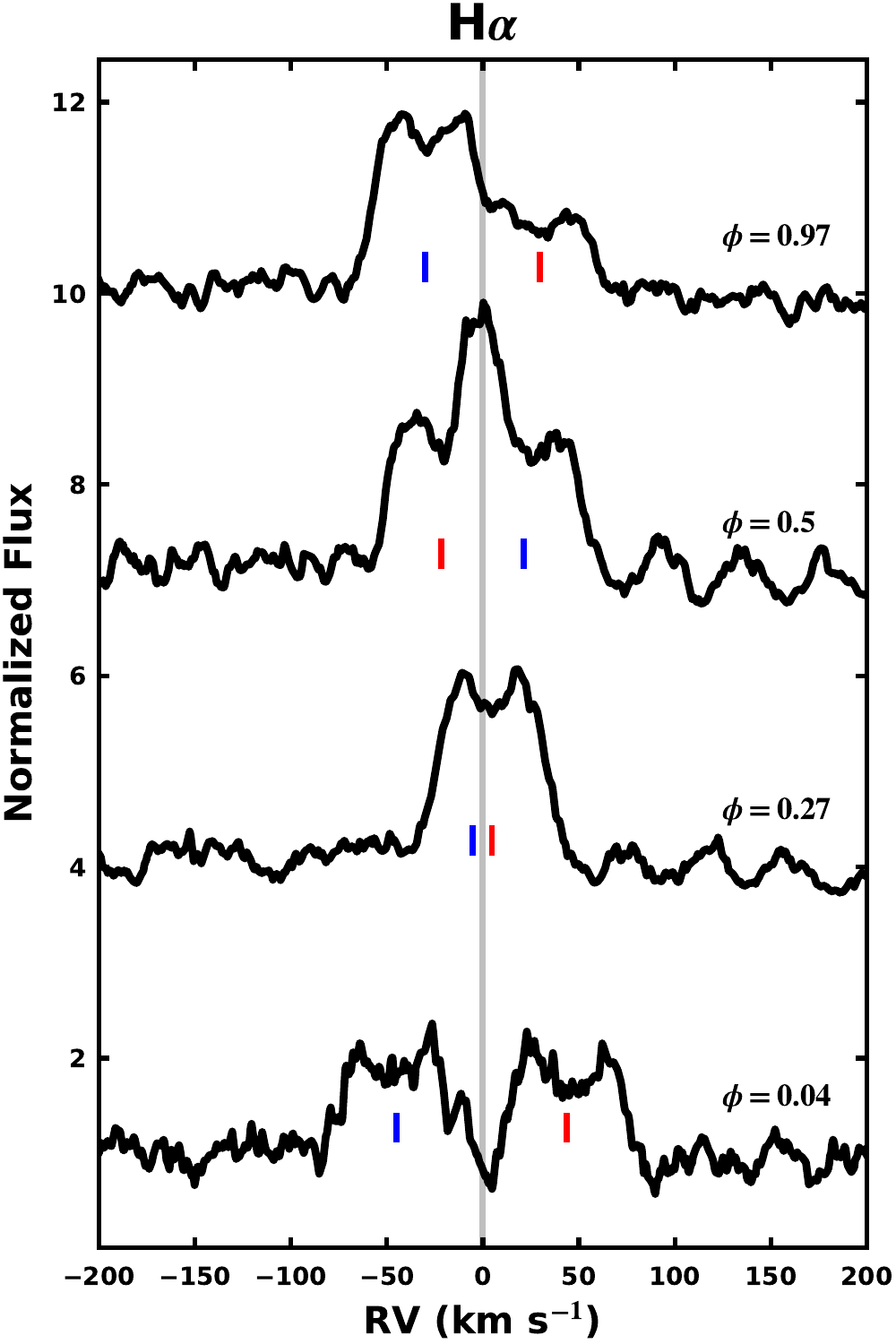}
\caption{H$\alpha$ line profiles for select SALT--HRS epochs. Spectra have been shifted to the system's center-of-mass velocity. Blue and red dashes mark the primary and secondary velocities, respectively. The orbital phase of each observation is provided to the right of each spectrum. }
\label{fig:ha}
\end{center}
\end{figure}

In this subsection we highlight the characteristics of two spectral features that trace stellar youth. 

{\bf H$\alpha$:} Chromospheric emission traces magnetic activity \citep[e.g.,][]{Skumanich1972}, which declines as stars age and spin down via magnetic breaking \citep{Weber&Davis1967}. The spread in late-M dwarf chromospheric activity, as probed by \hal\ in young clusters, is too large to determine a precise age \citep{Douglasetal2014,Krausetal2014,Fangetal2018Ha}. The timescale to observe M dwarf activity evolution is on the order of Gyrs \citep{Newtonetal2016,Newtonetal2017}. The presence of a close binary companion will also complicate a star's rotational evolution. Still, the presence of strong emission in this system, which is not particularly rapidly rotating, is consistent with youth. 

Figure \ref{fig:ha} presents four \hal\ epochs. The orbital phase is provided to the right of each curve, and the primary and secondary velocities are shown in the blue and red vertical dashes, respectively. The \hal\ line profile for each star is double peaked, characteristic of self-absorbed chromospheric emission \citep[e.g.,][]{Houdebineetal2012}. The strength of each component is variable, as highlighted by the comparison of the top and bottom epoch, the former of which may have been observed during a flaring event on the primary star. There are only three epochs where the \hal\ line profiles are fully separated. From these we compute average equivalent widths through numerical integration, finding $-2.2\pm0.3$ and $-2.1\pm0.4$ \AA\ for the primary and secondary, respectively, where the uncertainty is the standard deviation of the three measurements. These values are corrected for the diluting effect of the two continuum sources; for an average flux ratio of unity, this amounts to a factor of 2 increase.

{\bf Li:} The presence of Li in a stellar atmosphere can provide a powerful probe of stellar age as the element is rapidly burned at the base of the convective zone. For M 4.5 stars, like TOI 450, lithium supplies are exhausted between 20 and 45 Myr (\citealt{Mentuchetal2008}, using \citealt{Baraffeetal1998} models, and empirically, e.g., \citealt{Krausetal2014}). We do not detect the \lii\ 6708 \AA\ absorption line, consistent with our expectation for an M4.5 dwarf in the Columba association.  

\begin{figure}[t!]
\begin{center}
\includegraphics[width=0.47\textwidth]{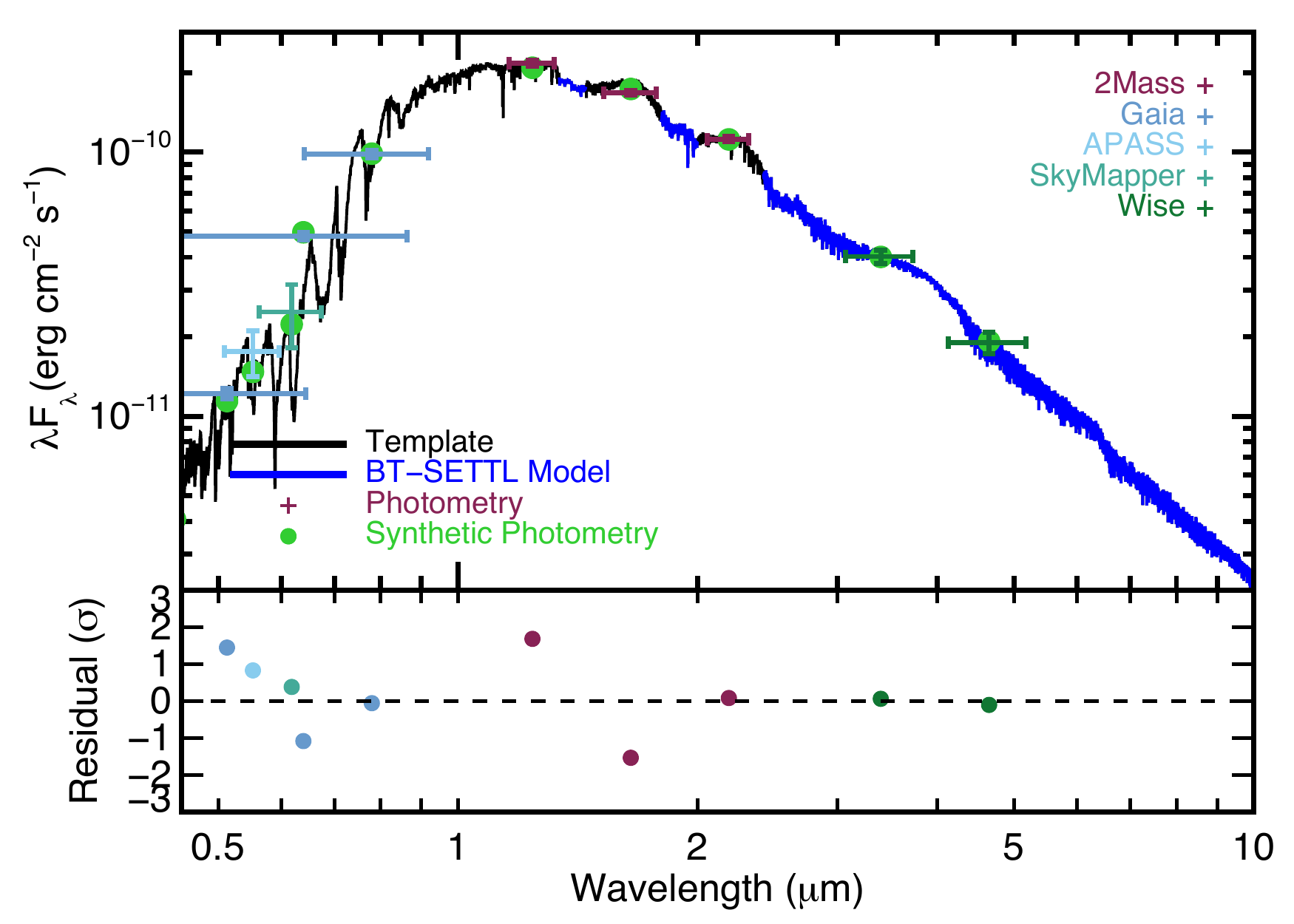}
\caption{Example of our spectral energy distribution (SED)-fitting procedure. The top subpanel shows an example template (black) and BT-SETTL model (blue) along the synthetic photometry (green). The observed photometry is colored by the source, with vertical errors indicating measurement errors and horizontal errors indicating the filter width. The bottom panel shows the residuals in units of standard deviations. The final residuals include errors from filter zero-points and hence do not perfectly match the comparison in the top panel. }
\label{fig:SED}
\end{center}
\end{figure}

\subsection{Quantities Derived from Unresolved Photometry}
\label{sed}

We fit the unresolved photometry assuming a single star following the method outlined in \citet{Mann2015b}. To briefly summarize, we compared unresolved photometry to a grid of optical and near-IR (NIR) spectral templates from \citet{Rayner2009} and \citet{Gaidos2014}. We use BT-SETTL models to fill in gaps in the spectra (e.g., past 2.4 $\mu$m). The free parameters are template selection, model selection, and three free parameters to handle systematic errors in the flux calibration and scaling between the spectra and photometry. We generate synthetic photometry from the templates using the appropriate filter profile. For our comparison, we use photometry from \gaia\ EDR3, the AAVSO Photometric All Sky Survey \citep{Hendenetal2015}, the SkyMapper survey \citep{Wolfetal2018}, the Two-Micron All-sky Survey \citep[2MASS;][]{Skrutskie2006}, and the Wide-field Infrared Survey Explorer \citep[ALLWISE;][]{Cutrietal2013}. We integrate the full spectrum to determine the bolometric flux (\fbol).

The fit yields an \fbol\ of $0.024\pm0.002\times 10^{-8}$\,erg\,cm$^{-2}$s$^{-1}$ and \teff\ of 3150$\pm$80\,K (determined from the assigned templates). The best-fit template spectra are all M4V--M5V, in good agreement with the CARMENES empirical-template match to our high-resolution spectra (Section \ref{rvs}). The final uncertainties account for both measurement errors and systematics in filter zero-points. We show an example fit in Figure~\ref{fig:SED}.

\section{Columba Membership}
\label{ff-columba}

\begin{figure*}[!t]
\begin{center}
\includegraphics[width=0.48\textwidth]{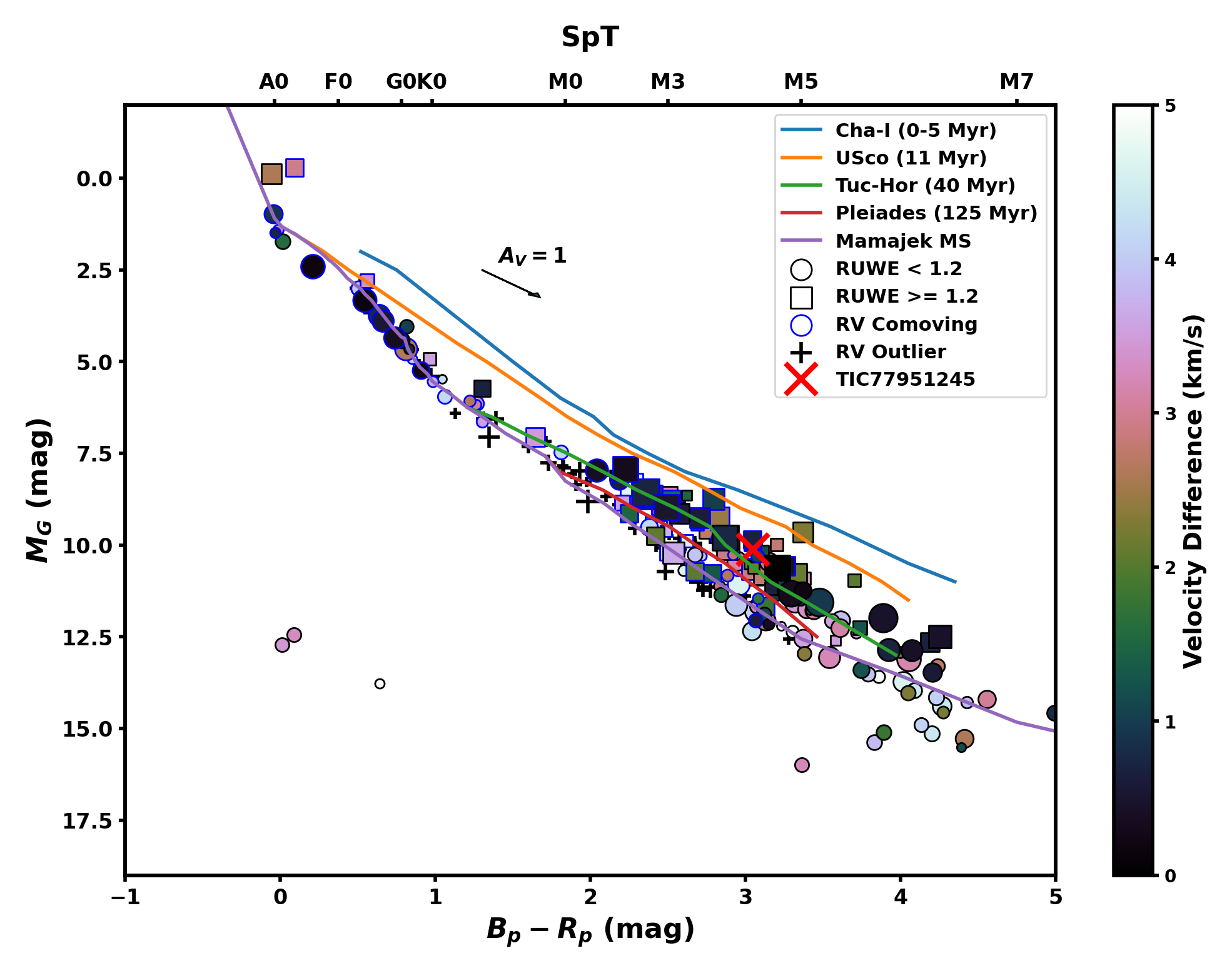}
\includegraphics[width=0.48\textwidth]{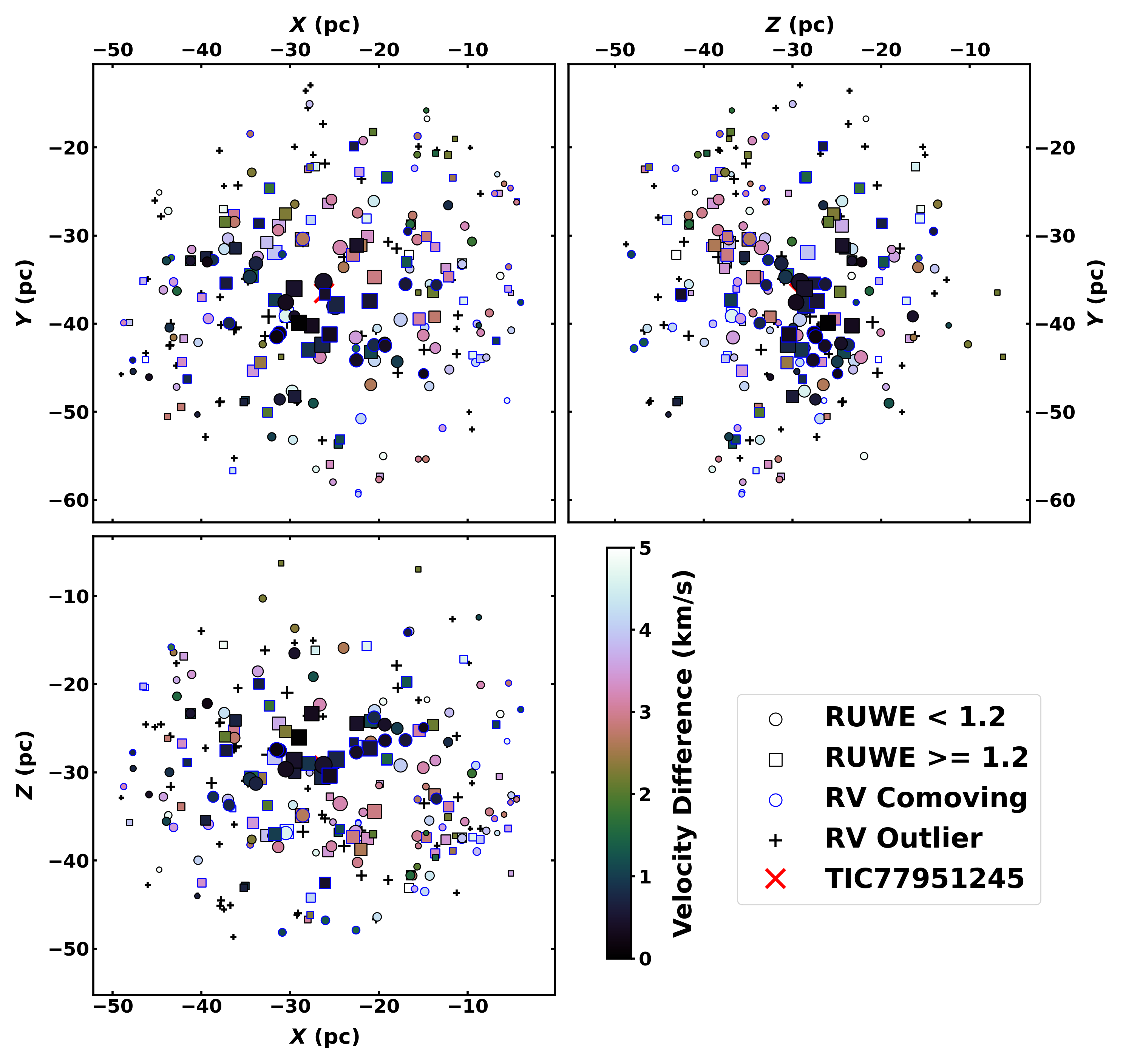}
\caption{{\tt FriendFinder} results for TOI 450, which recovers the Columba association. In each panel TOI 450 is labeled with a red $\times$. ``Friends" are plotted in circle if their \gaia\ RUWE is less than 1.2 (presumed single) and in squares if their \gaia\ RUWE is greater than 1.2 (presumed binary). The size of the point encodes its 3D distance from TOI 450 (larger is closer). The color encodes the tangential velocity difference from TOI 450 as shown in the color bars. {\bf Left}: Sky map of TOI 450 friends. {\bf Right}: XYZ spatial distributions of TOI 450 friends. \label{fig:FF-coord}}
\end{center}
\end{figure*}

\begin{figure}[!h]
\begin{center}
\includegraphics[width=0.48\textwidth]{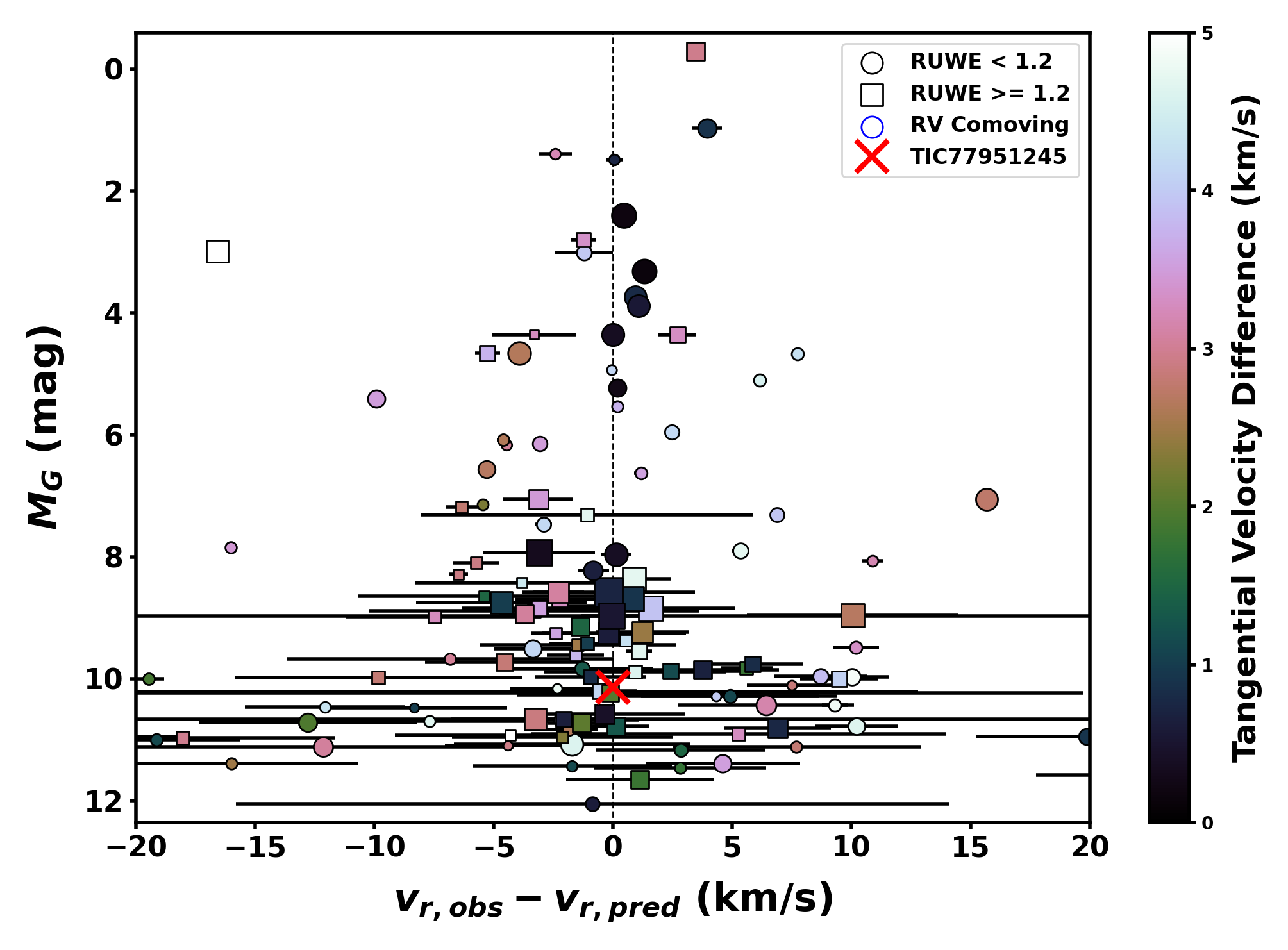}
\caption{Radial velocity difference friends from TOI 450 as a function of the absolute $G$ magnitude. Velocities compiled from \gaia\ EDR3 and the literature. The shape, size, and color coding scheme are described in Figure \ref{fig:FF-coord}.}
\label{fig:FF-rvs}
\end{center}
\end{figure}

\begin{figure}[!h]
\begin{center}
\includegraphics[width=0.48\textwidth]{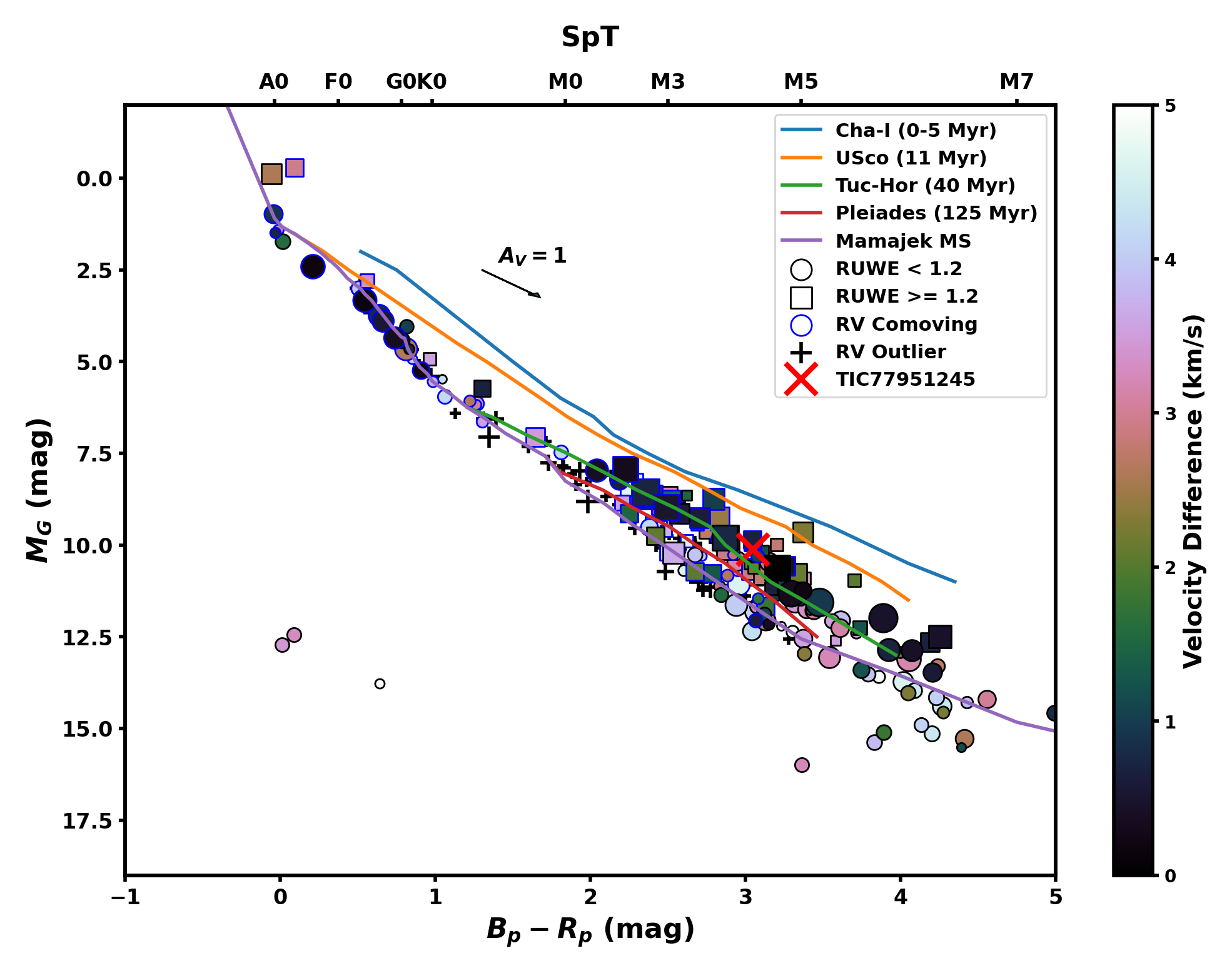}
\caption{\gaia\ color-magnitude diagram (CMD) of TOI 450 friends. The CMD distribution is broadly consistent the with Tuc-Hor empirical sequence at $\sim$40 Myr. The shape, size, and color coding scheme are described in Figure \ref{fig:FF-coord}.}
\label{fig:FF-cmd}
\end{center}
\end{figure}

TOI 450 was first proposed to be a candidate member of the Columba association by \citet{Gagne&Faherty2018}, who used Banyan-$\Sigma$ \citep{Gagneetal2018} to evaluate the five-dimensional kinematics of all stars within $D < 100$ pc and check for agreement with the pre-defined six-dimensional loci of the major known moving groups. (Banyan-$\Sigma$ predicts a 99.9\% Columba membership.)  \citet{CantoMartinsetal2020} subsequently measured a photometric rotational period of $P_{rot} = 5$ days, which, while on the upper envelope of the rotational sequence at $\tau \la 100$ Myr \citep[e.g.,][]{Rebulletal2016}, and is on the short-period end of typical mid-M field stars \citep{Newtonetal2016}. Our photometric analysis now shows that the stars are indeed substantially inflated over the MS (Table \ref{tab:fits}), implying that they are indeed young and still contracting to the MS. However, a precise age would substantially increase the value of TOI 450 in testing stellar evolutionary models, and the nature and age of Columba has remained unclear.

The Columba association was first identified as a subgroup within the notional ``Great Austral Young Association'' \citep{TorresCetal2001}, a conglomeration of the Tuc-Hor, Carina, and Columba associations \citep{Zuckermaneteal2001,TorresCetal2003,TorresCetal2006}. However, Columba was recognized to be more diffuse than many other associations \citep{TorresCetal2008}, which led to lower membership probabilities and a broader scope for incorporating additional members. This led to the addition of such far-flung systems as HD 984, HR 8799, and Kappa Andromedae to its census \citep[e.g.,][]{Zuckermanetal2011}, further loosening its definition and raising the probability that field contaminants and even other young associations were incorporated into its definition. With this in mind, a sample of 50 Columba members was used to fit an isochronal age of $42^{+6}_{-4}$ Myr \citep{Belletal2015}. The Gaia era now offers a new opportunity to revisit the definition of the Columba association, especially in providing a contextual age for TOI 450.

To identify candidate comoving neighbors (hereafter ``friends'') to TOI 450, we have used the software routine {\tt FriendFinder} \citep{Tofflemireetal2021} that is distributed in the {\tt Comove} package \footnote{\url{https://github.com/adamkraus/Comove}}. The {\tt FriendFinder} is a quicklook utility that adopts the Gaia astrometry and a user-defined RV (Table \ref{tab:rv}; $v_{rad} = 23.74$ km/s) for a given science target, computes the corresponding XYZ space position and UVW space velocity, and then screens every Gaia source within a user-defined 3D radius ($R = 25$ pc) to determine if its sky-plane tangential velocity matches the (re-projected) value expected for comovement within a user-defined threshold ($\Delta v_{tan} \le 5$ km/s). Plots are then generated for the friends' sky-plane positions, $UVW$ velocities, and RV distribution (using Gaia RVs and any others that we manually add). Finally, additional catalogs are also queried to produce plots of the friends' GALEX UV photometric sequence \citep{Bianchietal2017} normalized by their 2MASS $J$-band flux, and WISE infrared photometric color sequence.

In Figure \ref{fig:FF-coord} (left), we plot a sky map of the 467 Gaia sources that were selected as friends. Each source's offset in $v_{tan}$ is shown with its shading, from dark ($\Delta v_{tan} =0$ km/s) to light ($\Delta v_{tan} = 5$ km/s), and the 3D distance is shown with its size. Sources with $RUWE > 1.2$ (denoting potential binarity) are shown with squares, while others are shown with circles. If a source also has a known RV, then the point is outlined in blue if the RV also agrees with comovement to within $\Delta v_{rad} < 5$ km/s, whereas objects with discrepant RVs are replaced with crosses. Visual inspection shows that there is an overdensity of large, dark points surrounding TOI 450, elongated into an ovoid that is aligned roughly N-S. Many of these sources are also comoving in RV, and hence in their full three-dimensional velocity vector, so we conclude that there is likely a coherent comoving population around TOI 450.

In Figure \ref{fig:FF-coord} (right) we also show the XYZ spatial distribution of the full sample of friends. The locus of large dark points (denoting the apparently young, comoving stellar population) is concentrated in the center around TOI 450, with an approximate full extent of $\pm 30$ pc in X, $\pm 15$ pc in Y, and $\pm 10$ pc in Z. We note that there does appear to be potentially coherent structure among stars that are not as clearly comoving, especially for the pink points ($\Delta v_{tan} \sim 3$ km/s) that fall at $+Y$ and $-Z$ from the central locus. Those near-comoving and nearly-cospatial sources include stars that have been identified as potential Tuc-Hor members, further hinting at the existence of a kinematic link (but not an identical nature) between Columba and Tuc-Hor.

In Figure \ref{fig:FF-rvs}, we plot the corresponding distribution of $\Delta v_{rad}$ for all friends that have known RV measurements in Gaia or in other catalogs. There is again a notable excess of sources that are comoving with TOI 450 to within $\Delta v_{rad} < 5$ km/s; the velocity distribution of the thin disk is much larger ($\sigma_{vrad} \sim 30$ km/s; ref), so an overdensity on a scale of 5 km/s further emphasizes the likely existence of a coherent comoving stellar population.

Finally, in Figure \ref{fig:FF-cmd}, we show the ($M_G$, $Bp-Rp$) color-magnitude diagram (CMD) for all friends that have valid photometry in all bands. The CMD further demonstrates that TOI 450 is not merely surrounded by a comoving population, but that it is relatively young; the large dark points form a notable pre-MS that approximately traces a reference sequence for Tuc-Hor \citep{Krausetal2014}. The presence of numerous sources along the field MS indicates that the friend population is substantially contaminated with field interlopers, and hence can not simply be adopted for further demographic studies. However, there is an apparent pre-MS turn-on at $M_G \sim 8$ mag;
most sources above this limit have Gaia RVs that can be used to reject field interlopers, while the sources fainter than this limit can be screened by requiring them to fall above the visually obvious divide separating the pre-MS and MS sequences. The existence of a coherent pre-MS population demonstrates that the coherent comoving stellar population is likely young, agreeing with the apparent young age of TOI 450.

The {\tt FriendFinder} also outputs plots of the GALEX NUV flux normalized by the 2MASS $J$-band flux and the WISE W1-W3 color, both as a function of the \gaia\ color. We exclude these plots for brevity, but both sequences behave as expected for a 40 Myr sequence. The GALEX plot shows a sequence sitting above the older and less active Hyades, while the WISE plot shows no evidence for infrared excesses (i.e., candidate members are not disk bearing).

In summary, the evidence strongly indicates that TOI 450 is embedded in a comoving and cospatial young stellar population that we recover as the Columba association. A full analysis of its age and demographics is beyond the scope of this current effort, and the age of TOI 450 could be further clarified with dedicated studies of lithium depletion and rotational spindown in the population. Because of Columba's complicated membership history, we do not adopt a previously published age, however, given the broad consistency between this population's CMD sequence (Figure \ref{fig:FF-cmd}) and the isochronal sequence of Tuc-Hor, it seems broadly warranted to adopt a similar age of $\tau \sim 40$ Myr \citep[e.g.,][]{Krausetal2014} for TOI 450 and its host population.

\section{Eclipsing Binary Fit}
\label{ebfit}

To derive the fundamental parameters of the TOI 450 binary system, we jointly fit the RV measurements and eclipse light curves with a modified version of the {\tt misttborn} code \citep{Mannetal2016a, Johnsonetal2018}. The RVs are described by a Keplerian orbit, and eclipses are modeled with the analytic transit code {\tt batman} \citep{Kreidberg2015}, diluted by the companion's secondary light, assuming a quadratic limb-darkening law \citep{DiazCordovesetal1992}. Both data sets are fit within a Markov Chain Monte Carlo (MCMC) framework using {\tt emcee} \citep{Foreman-Mackeyetal2013}. The model has 23 parameters: the time of periastron passage ($T_0$), orbital period ($P$), semi-major axis divided by the sum of the stellar radii ($a/(R_1+R_2)$), the ratio of the stellar radii ($R_2/R_1$), cosine of the orbital inclination ($\cos{i}$), mass ratio ($q$), sum of the velocity semi-major amplitudes ($K_1+K_2$), center of mass velocity ($\gamma$), and a zero-point offset between SALT--HRS and HARPS RVs ($\mu$). The orbital eccentricity ($e$) and the longitude of periastron ($\omega$) are fit with the combined parameterization of $\sqrt{e}\sin{\omega}$ and $\sqrt{e}\cos{\omega}$, which is computationally efficient and avoids biases at low and high eccentricities inherent in other approaches \citep[e.g.,][]{Eastmanetal2013}. Finally, for each eclipse light-curve filter ($r_p$, \tess, $I$) there are four parameters: a central surface-brightness ratio ($J_2/J_1$), two quadratic limb-darkening coefficients (LDCs; $q_1$, $q_2$), and a photometric jitter term ($\sigma_{LC}$). The $q_1$ and $q_2$ LDCs are the \citet{Kipping2013} triangular sampling parameterization of the standard quadratic LDCs $u_1$ and $u_2$, where $q_1 = (u_1 + u_2)^2$ and $q_2 = u_1/2(u_1+u_2)$. Given their similarity, we assume the primary and secondary have the same LDCs. With the exception of the photometric jitter terms, which are explored in logarithmic space, all parameters are explored in linear space. 

We fit detrended light curves in this approach. For \tess, we use the Gaussian process model in Section \ref{rotp} to remove stellar variability. To reduce computation time, we only fit the \tess\ light curve in 1.1 day windows centered on the superior and inferior conjunctions (determined from initial $e$ and $\omega$ values from an orbit fit to the RV measurements). For the LCO $r'$- and $I$-band light curves, we fit a line to the out-of-eclipse regions, which is appropriate for the timescale of variability we observe in the \tess\ light curve, and normalize the light curve with that fit. 

Certain choices in the measurements that are fit are made to reduce the effect of systematic and/or correlated measurement errors. Similarly, choices in the fit parameters themselves are made to reduce covariance between fit parameters. For the stellar RVs, we fit the primary RV ($RV_1$) and the difference between the primary and secondary RV ($RV_1-RV_2$) in order to reduce the effect of correlated RV errors due to epoch-dependent shifts in the wavelength calibration (i.e., correlated shifts in $RV_1$ and $RV_2$). Fitting the RV difference also reduces the fit dependence on the zero-point difference between the SALT--HRS and HARPS instruments. For the fit parameters, we elect to fit the sum of velocity semi-major amplitudes ($K_1+K_2$) and the mass ratio ($q$), as opposed to $K_1$ and $K_2$, to reduce the covariance between these parameters and the center-of-mass velocity. 

Our analysis assumes that gravitational darkening, ellipsoidal variations, reflected light, and light travel time corrections are all negligible.  We confirm this by creating a model with our best-fit values using the {\tt eb}\footnote{https://github.com/mdwarfgeek/eb} \citep{Irwinetal2011} package (a C and python implementation of the well-established Nelson–Davis–Etzel binary model used in the EBOP code and its variants; \citealt{Etzel1981,Popper&Etzel1981}), finding the deviations from our simplified model are a factor of $\sim$30 smaller than the uncertainty of our highest-precision photometric data set ($r^\prime$), and a factor of $\sim$40 smaller than our radial velocity precision. The most significant astrophysical ingredient missing from our model is the effect of star spots, which we address in Section \ref{spots}.

Table \ref{tab:fits} lists our model's fit parameters and their associated priors. In general, the bounds provided by our uniform priors ($\mathcal{U}$) do not influence the parameter exploration but are listed for transparency. The only exceptions are the uniform priors on $q_1$ and $q_2$, which bound the physical parameter space of the LDCs. Although it is common practice to subject the exploration of $q_1$ and $q_2$ to Gaussian priors on the true quadratic LDCs ($u_1$, $u_2)$ based on predictions of their filter specific values (e.g, \citet{Claret&Bloemen2011}; \cite{Claret2017}), recent work by \citet{Patel&Espinoza2022} has shown systematic offsets in theoretical predictions and empirically derived LDC values that are especially large for cool stars ($\Delta u_{1,2} \geq 0.25$). For this reason, we do not place priors on the derived $u_1$ and $u_2$ values. The remaining priors on the radius ratio and central surface-brightness ratios are described in the following section.

\subsection{Priors Informed by Spectroscopic Analysis}
\label{priors}

\begin{figure}[t!]
\begin{center}
\includegraphics[width=0.47\textwidth]{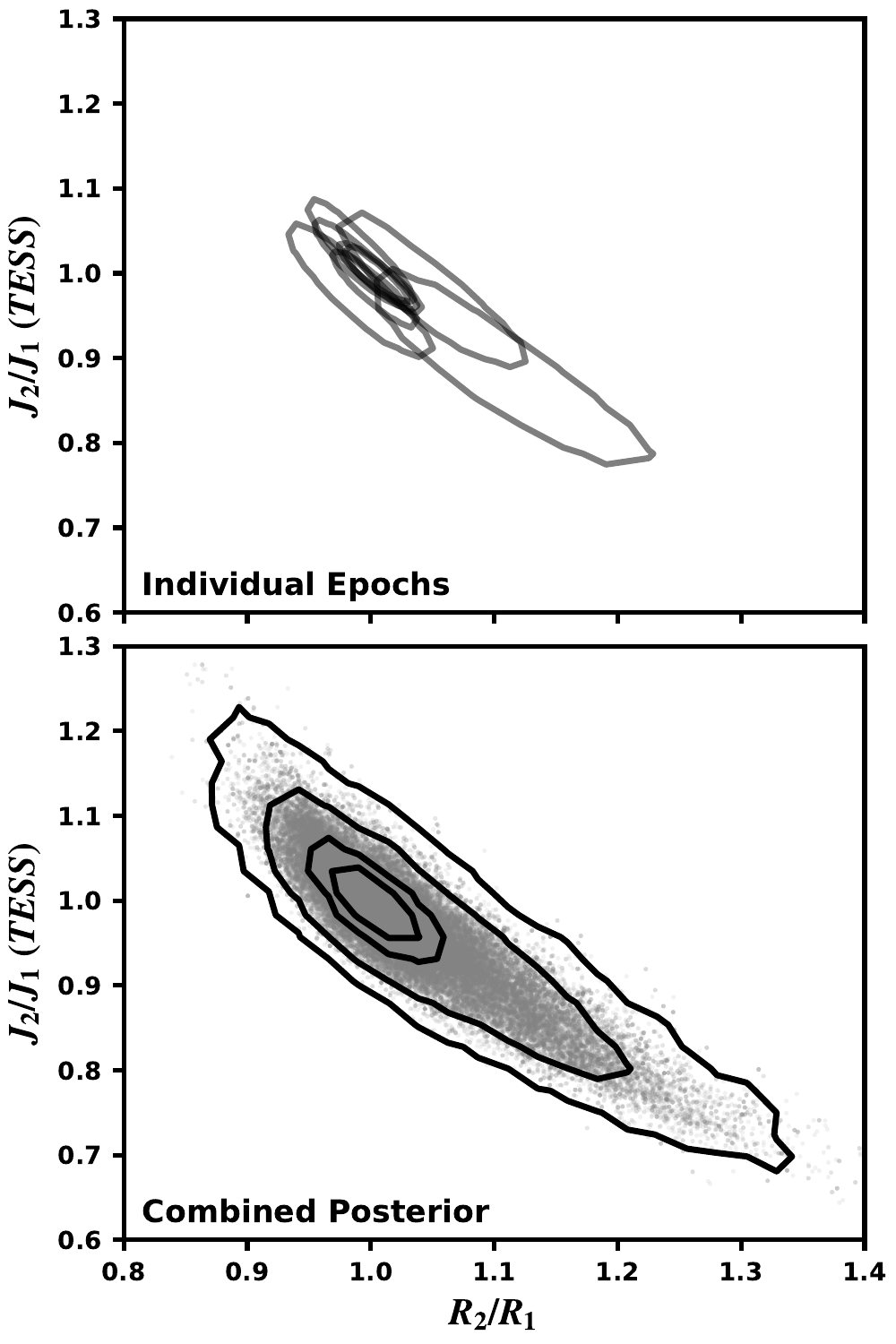}
\caption{Radius ratio and surface-brightness ratio posterior for the \tess\ bandpass from our fit to the SALT--HRS flux ratios and broadband spectral energy distribution. The top panel presents the 68\% contours for the eight individual epochs. The bottom panel is the joint posterior with contours enclosing the 50th, 68th, 95th, and 99.7th percentiles. }
\label{fig:kde}
\end{center}
\end{figure}

In a traditional, double-eclipsing, EB system, the combination of the primary and secondary eclipse is sufficient to constrain the central surface-brightness ratio ($J_2/J_1$) and radius ratio ($R_2/R_1$), such that an informed prior on either is not strictly required. Even so, constraints from spectroscopy have been used in many previous analyses \citep[e.g.,][]{Stassunetal2006}. Recent work has shown that these parameters can be independently constrained to a greater degree with measurements of the wavelength-dependent stellar flux ratio from high-resolution spectra and/or joint spectral energy distribution (SED) fitting \citep[e.g.,][]{Krausetal2017,Torresetal2019,Gillenetal2020}. The impact of spectroscopic constraints is far reaching as they directly affect other fit parameters ($a/(R_1+R_2)$, $\cos{i}$) and derived quantities ($M_1$, $M_2$, $a$, $R_1$, $R_2$). In the case of TOI 450, its single grazing eclipse necessitates an informed prior in order to perform a meaningful fit to the system. In practice, the lack of a secondary eclipse does limit the inclination such that measurements of the stellar radii can be made with $\sim$30\% precision. However, this is insufficient to rigorously test stellar evolution models, and ignores valuable information contained in our spectra. In this section we describe the construction of a joint surface-brightness ratio-radius ratio prior. 

With the wavelength-dependent optical flux ratios measured from the SALT--HRS spectra (Section \ref{FR}) and the compiled broadband optical and NIR photometry (Section \ref{sed}), we fit the combination of two synthetic stellar templates from the BT-SETTL atmospheric models \citep{Allard2013} within an MCMC framework using \texttt{emcee}. We restrict our comparison to solar metallicity models and a surface gravity \logg\ of 5. We test other surface gravities and find the effect is negligible. Thus, \teff\ uniquely determines the model selection. 

The six free parameters are the primary \teff\ ($T_P$), the companion \teff\ ($T_C$), a scale factor for each star ($S1$ and $S2$), and two parameters that describe underestimated uncertainties in the unresolved photometry ($s_1$ [mags]) and the spectroscopic flux ratios ($s_2$ fractional). The scale factors describe the ratio of the measured flux to that of the model. 

For each step in the MCMC, we scale and combine the two model spectra to form an unresolved spectrum. We convolve this spectrum with the relevant filter profiles \citep[e.g.,][]{Cohen2003, Mann2015a}, which we compare directly with the observed SED photometry (10 photometric bands). We also compute the spectroscopic flux ratio in optical bands matching the output from Section \ref{FR} (30 orders). Constraints from the SED and flux ratios are weighted equally in the likelihood function, assuming Gaussian errors after adding in the $s$ parameters in quadrature with measurement errors. 

The MCMC explores the scale factors using log-uniform priors, and all other parameters using linear-uniform priors. We run the fit with 20 walkers for 10,000 steps following a burn-in of 2000 steps. This is more than sufficient for convergence based on the autocorrelation time. 

The atmospheric models likely have systematic errors due to missing opacities \citep{Mann2013c}. However, the effect is almost identical on both stellar components due to a common model grid and similar temperatures. We also mitigate this effect by shifting our posteriors into parameter ratios. Specifically, we convert the posteriors on $T_P$ and $T_C$ into the corresponding surface-brightness ratios in the $r'$, \tess, and $I$ bandpasses using the same BT-SETTL models and the posteriors. For radius, we use the scale factors, which are proportional to $R^2/D^2$. The two-component stars are the same distance, which makes it trivial to convert the ratio of the scale factors to the radius ratio. 

We perform our fit for each of the eight SALT--HRS epochs where the stellar velocity separation is large enough for robust flux-ratio determinations. This is preferable to fitting the average of the eight because they span a range of rotational phases (Figure \ref{fig:FR_epochs}) allowing for the range of flux ratios presented by the system. Joining the posteriors of the derived parameters, $J_2/J_1$ and $R_2/R_1$, we create a Gaussian kernel-density estimate (KDE) for each filter ($r'$, \tess, and $I$), which serves as the priors for our eclipse model. Figure \ref{fig:kde} presents the 68\% contours of the \tess-specific posteriors for individual epochs (top panel) and a contour plot of combined posterior from which we compute a Gaussian KDE (bottom panel). The LCO $r^\prime$ and $I$-band versions follow the same basic shape, centering at a radius ratio and surface-brightness ratio of 1.

\subsection{Results}
\label{results}

\begin{figure}[t!]
\begin{center}
\includegraphics[width=0.48\textwidth]{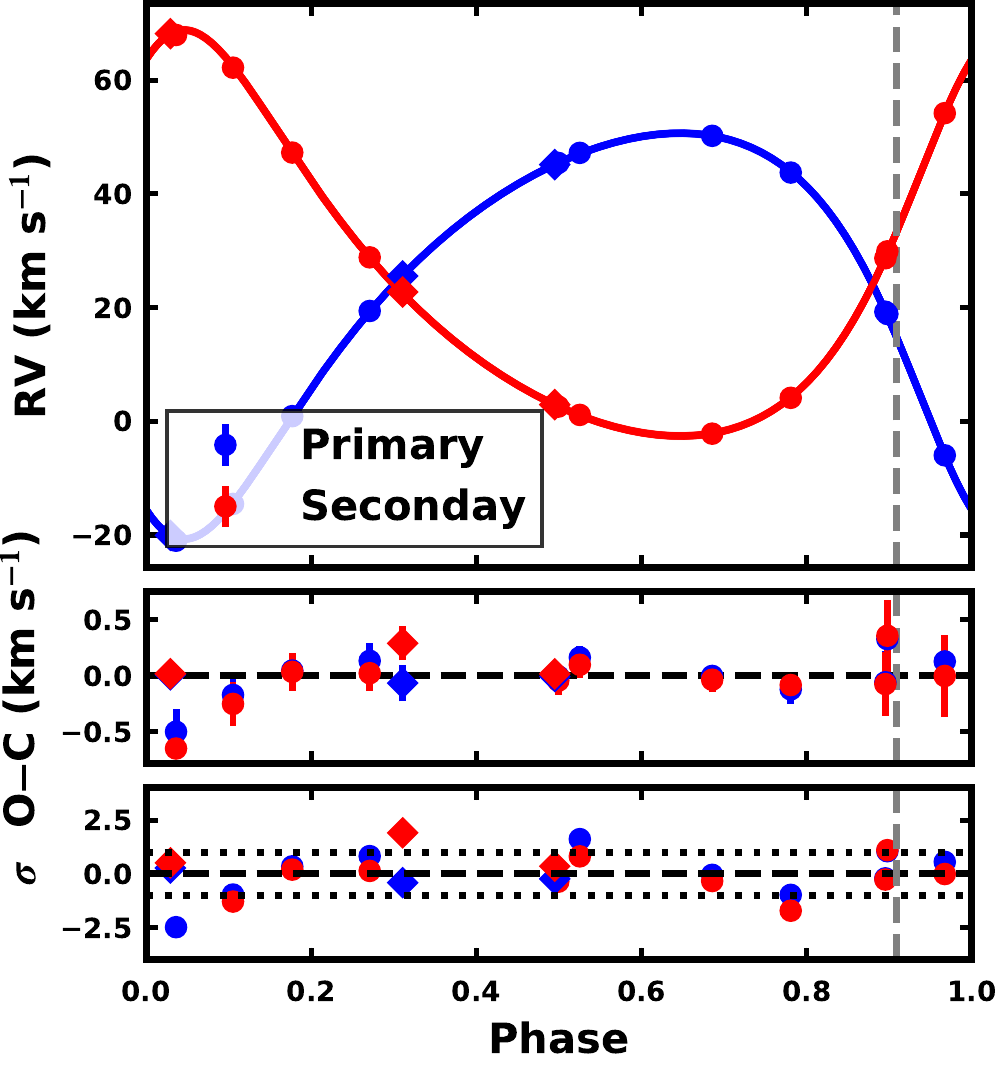}
\includegraphics[width=0.48\textwidth]{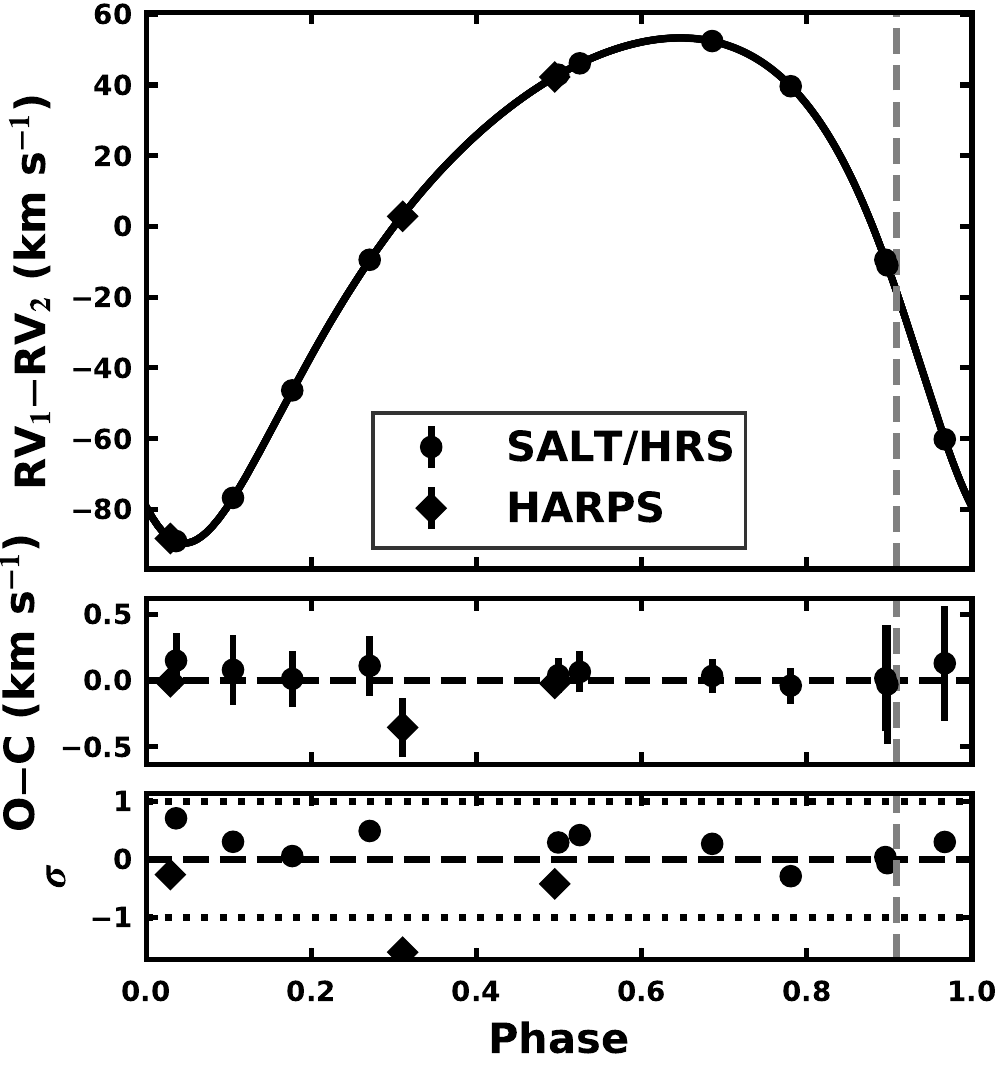}
\caption{Radial velocity orbital solution for TOI 450. The top panel set presents the radial velocities for the primary (blue) and secondary (red). The bottom panel set presents the RV difference. In each panel set, SALT--HRS and HARPS data are shown with circles and diamonds, respectively. The best-fitting orbital solution from our combined fit is shown as the solid lines. The bottom panels in each set present the fit residuals in \kms\ and in units of their measurement error ($\sigma$). The gray vertical line marks the phase of the primary eclipse. Phase $\phi = 0$ corresponds to periastron passage. }
\label{fig:rvs}
\end{center}
\end{figure}

\begin{figure}[t!]
\begin{center}
\includegraphics[width=0.47\textwidth]{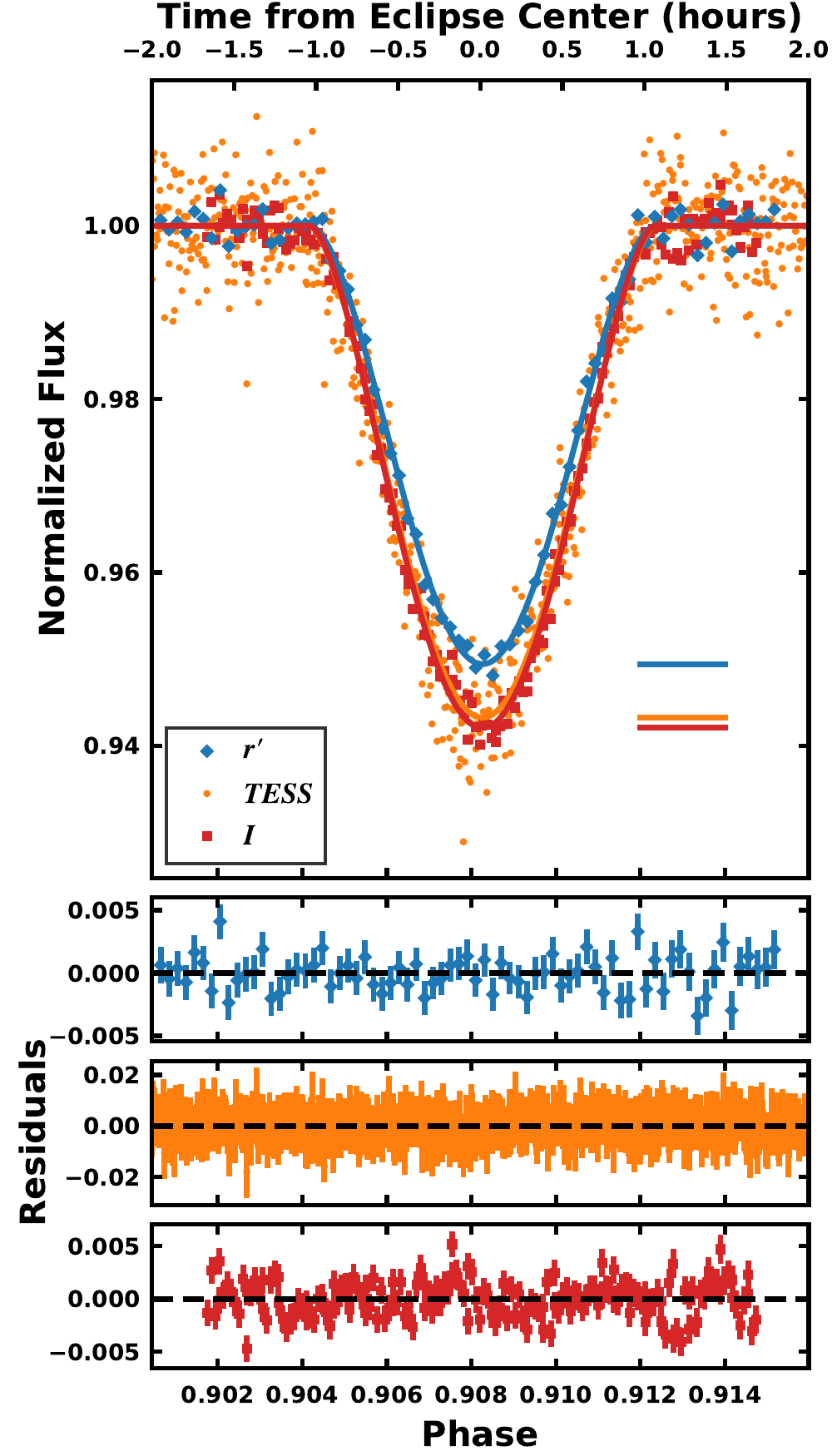}
\caption{Phase-folded eclipse light curves in the $r'$ (blue diamonds), \tess\ (orange circles), and $I$ (red squares) bandpasses. The best-fitting eclipse model from our combined fit is over-plotted in each filter in its associated color. Horizontal lines at the right of the plot highlight the eclipse depth in each filter. A strong color dependence is observed due to wavelength-dependent limb-darkening.}

\label{fig:eclipse}
\end{center}
\end{figure}
\begin{deluxetable*}{l c c c c c}[!h]
\tablecaption{Unspotted Fit Variations
\label{tab:fits}}
\tablewidth{0pt}
\tabletypesize{\footnotesize}
\tablecolumns{6}
\phd
\tablehead{
  \colhead{Parameter} &
  \colhead{Prior} & 
  \colhead{$r_p$} &
  \colhead{\tess} &
  \colhead{$I$} &
  \colhead{Combined} 
}
\startdata 
\multicolumn{6}{l}{\bf Fit parameters} \\
$T_0$ (BJD-2457000) & ...                      & $1433.437 \pm 0.003$      & $1433.436 \pm 0.003$         & $1433.437 \pm 0.003$         & $1433.437 \pm 0.003$     \\
$P$ (days) & ...                               & $10.714767 \pm 0.000006^a$& $10.714767 \pm 0.000006$     & $10.714767 \pm 0.000006^a$   & $10.714762 \pm 0.000005$ \\
$a/(R_1+R_2)$ & $\mathcal{U}(10,60)$           & $20.5^{+0.3}_{-0.4}$      & $20.3 \pm 0.3$               & $20.7 \pm 0.2$               & $20.6 \pm 0.2$           \\
$R_2/R_1$ & RR--SBR KDE                        & $1.01^{+0.04}_{-0.03}$    & $1.01^{+0.05}_{-0.03}$       & $1.01^{+0.05}_{-0.03}$       & $1.00 \pm 0.02$          \\
$\cos{i}$ & $\mathcal{U}(0,1)$                 & $0.048 \pm 0.001$         & $0.0477^{+0.0011}_{-0.0010}$ & $0.0471^{+0.0007}_{-0.0006}$ & $0.0473 \pm 0.0004$      \\
\\
$q_{1, r'}$ & $\mathcal{U}(0,1)$               & $0.6 \pm 0.3$             & ...                          & ...                          & $0.7 \pm 0.2$            \\
$q_{2, r'}$ & $\mathcal{U}(0,1)$               & $0.5 \pm 0.3$             & ...                          & ...                          & $0.7 \pm 0.2$            \\
$J_2/J_1 (r')$ & RR--SBR KDE                   & $0.99^{+0.07}_{-0.08}$    & ...                          & ...                          & $1.01 \pm 0.04$          \\
ln $\sigma_{LC, r'}$ & $\mathcal{U}(-12,1)$    & $-10 \pm 2$               & ...                          & ...                          & $-10 \pm 2$              \\
\\
$q_{1, \tess}$ & $\mathcal{U}(0,1)$            & ...                       & $0.5 \pm 0.3$                & ...                          & $0.4 \pm 0.2$            \\
$q_{2, \tess}$ & $\mathcal{U}(0,1)$            & ...                       & $0.5 \pm 0.3$                & ...                          & $0.5 \pm 0.3$            \\
$J_2/J_1$ (\tess)& RR--SBR KDE                 & ...                       & $0.99^{+0.04}_{-0.05}$       & ...                          & $1.00 \pm 0.03$          \\
ln $\sigma_{LC, \tess}$ & $\mathcal{U}(-12,1)$ & ...                       & $-10 \pm 1$                  & ...                          & $-10 \pm 1$              \\
\\
$q_{1, I}$ & $\mathcal{U}(0,1)$                & ...                       & ...                          & $0.5 \pm 0.3$                & $0.5^{+0.3}_{-0.2}$      \\
$q_{2, I}$ & $\mathcal{U}(0,1)$                & ...                       & ...                          & $0.3^{+0.3}_{-0.2}$          & $0.2^{+0.2}_{-0.1}$      \\
$J_2/J_1 (I)$ & RR--SBR KDE                    & ...                       & ...                          & $0.99^{+0.03}_{-0.04}$       & $1.00 \pm 0.02$          \\
ln $\sigma_{LC, I}$ & $\mathcal{U}(-12,1)$     & ...                       & ...                          & $-6.8 \pm 0.1$               & $-6.8 \pm 0.1$           \\
\\
$K_1+K_2$ (km/s) & $\mathcal{U}(25,100)$       & $71.43 \pm 0.04$          & $71.43 \pm 0.04$             & $71.43 \pm 0.04$             & $71.43 \pm 0.04$         \\
$q\ (M_2/M_1)$ & $\mathcal{U}(0.8,1.2)$        & $1.000^{+0.002}_{-0.001}$ & $1.000 \pm 0.001$            & $1.000 \pm 0.002$            & $1.000 \pm 0.001$        \\
$\sqrt{e}\sin{\omega}$ & $\mathcal{U}(-1,1)$   & $0.282 \pm 0.001$         & $0.282 \pm 0.001$            & $0.282 \pm 0.001$            & $0.282 \pm 0.001$        \\
$\sqrt{e}\cos{\omega}$ & $\mathcal{U}(-1,1)$   & $-0.4661 \pm 0.0006$      & $-0.4660 \pm 0.0006$         & $-0.4662 \pm 0.0006$         & $-0.4661 \pm 0.0006$     \\
$\gamma$ (km/s) & $\mathcal{U}(14,34)$         & $24.03 \pm 0.03$          & $24.03 \pm 0.03$             & $24.03 \pm 0.03$             & $24.03 \pm 0.03$         \\
$\mu$ (km/s) & ...                             & $0.29 \pm 0.05$           & $0.29 \pm 0.05$              & $0.29 \pm 0.05$              & $0.29 \pm 0.05$          \\
\hline
\multicolumn{6}{l}{\bf Derived Orbital Parameters} \\
$T_{p}$ (BJD-2457000)$^b$ &                          & $1432.4534 \pm 0.0002$    & $1432.4531 \pm 0.0003$       & $1432.4535 \pm 0.0001$       & $1432.4535 \pm 0.0001$   \\
$K_1$ (\kms)&                                  & $35.71 \pm 0.02$          & $35.71 \pm 0.02$             & $35.71 \pm 0.02$             & $35.71 \pm 0.02$         \\
$K_2$ (\kms)&                                  & $35.72 \pm 0.04$          & $35.72 \pm 0.04$             & $35.72 \pm 0.04$             & $35.72 \pm 0.04$         \\
$i$ (degrees)&                                 & $87.24^{+0.06}_{-0.07}$   & $87.27 \pm 0.06$             & $87.30 \pm 0.04$             & $87.29 \pm 0.02$         \\
$e$&                                           & $0.2968 \pm 0.0004$       & $0.2968 \pm 0.0004$          & $0.2969 \pm 0.0004$          & $0.2969 \pm 0.0004$      \\
$\omega$ (radian) &                            & $2.597 \pm 0.002$         & $2.597 \pm 0.002$            & $2.598 \pm 0.002$            & $2.597 \pm 0.002$        \\
$a$ ($au$) &                                   & $0.06726 \pm 0.00004$     & $0.06726 \pm 0.00004$        & $0.06725 \pm 0.00004$        & $0.06725 \pm 0.00004$    \\
\hline
\multicolumn{6}{l}{\bf Derived Stellar Parameters} \\
$M_1$ $(M_\odot)$&                             & $0.1768 \pm 0.0004$       & $0.1768 \pm 0.0004$          & $0.1768 \pm 0.0004$          & $0.1768 \pm 0.0004$      \\
$M_2$ $(M_\odot)$ &                            & $0.1768 \pm 0.0003$       & $0.1767 \pm 0.0003$          & $0.1767 \pm 0.0003$          & $0.1767 \pm 0.0003$      \\
$R_1 (R_\odot)$ &                              & $0.351 \pm 0.006$         & $0.354^{+0.007}_{-0.008}$    & $0.348^{+0.006}_{-0.008}$    & $0.351 \pm 0.003$        \\
$R_2 (R_\odot)$ &                              & $0.353^{+0.012}_{-0.010}$ & $0.357^{+0.012}_{-0.009}$    & $0.351^{+0.012}_{-0.008}$    & $0.351^{+0.005}_{-0.004}$\\
\hline
\multicolumn{6}{l}{\bf Derived Limb Darkening Parameters} \\
$u_{1, r'}$    &                               & $0.7^{+0.4}_{-0.5}$       & ...                          & ...                          & $1.2^{+0.2}_{-0.3}$      \\
$u_{2, r'}$    &                               & $-0.0^{+0.5}_{-0.4}$      & ...                          & ...                          & $-0.3 \pm 0.3$           \\
$u_{1, \tess}$ &                               & ...                       & $0.7^{+0.5}_{-0.4}$          & ...                          & $0.6 \pm 0.3$            \\
$u_{2, \tess}$ &                               & ...                       & $-0.0^{+0.5}_{-0.4}$         & ...                          & $-0.0^{+0.4}_{-0.3}$     \\
$u_{1, I}$     &                               & ...                       & ...                          & $0.4^{+0.4}_{-0.2}$          & $0.2 \pm 0.2$            \\
$u_{2, I}$     &                               & ...                       & ...                          & $0.3 \pm 0.4$                & $0.5^{+0.2}_{-0.3}$      \\
\enddata
\tablenotetext{a}{For these single-transit fits, a strict orbital period prior informed by the \tess-only fit is used to ensure a more direct comparison between the derived parameters from the fit variations.}
\tablenotetext{b}{Time of primary eclipse.}
\end{deluxetable*}

We perform our joint RV and light-curve fit for each photometric data set ($r'$, \tess, $I$) independently, which we call {\it individual} fits, and a final fit that combines all of the eclipse light curves, which we call the {\it combined} fit. Each fit employs 115 walkers where convergence is assessed following the scheme outlined in Section \ref{rotp}. In Table \ref{tab:fits} we provide the results of each fit parameter as well as some derived quantities. Values and their uncertainties are the posterior's median and central 68\% interval, respectively. We note that in order to more directly compare the results from the individual fits with single eclipses ($r^\prime$, $I$) to the \tess\ and combined fits, we place a strict Gaussian prior on the period for these two fits, informed by the period posterior from individual \tess\ fit. 

Figure \ref{fig:rvs} presents the RV orbital solution from our combined fit in the $RV_1$ and $RV_2$ (top panels), and $RV_1-RV_2$ (bottom panels) spaces, along with their residuals in \kms\ and in units of the measurement error ($\sigma$). RVs are presented as a function of the orbital phase where $\phi = 0$ corresponds to periastron passage. In the first $O-C$ panel of the $RV_1$, $RV_2$ panel set (top panels), specific SALT--HRS epochs show correlated errors where both the primary and secondary velocities are offset in the same direction from the best-fit model. Specifically the measurements at orbital phase, $\phi =$ 0.04, 0.11, and 0.50, highlight our motivation in fitting $RV_1$ and $RV_1-RV_2$, as opposed to $RV_1$ and $RV_2$.

Figure \ref{fig:eclipse} presents the $r^\prime$, \tess, and $I$ eclipse light curves with the combined fit model overlaid. Horizontal lines to the right show the eclipse depth in each filter. The residuals of each filter are also provided in the subsequent panels. Here the wavelength dependence of the eclipse depth is clear, where the shortest wavelength ($r^\prime$) has the shallowest depth. This behavior is expected for a grazing eclipse due to the wavelength dependence of limb-darkening. The same behavior could, in principle, result from specific spot patterns, which we discuss further in Section \ref{spots}. Here, given our limited prior knowledge of the LDCs, we find they are sufficiently flexible to describe the system's wavelength-dependent limb-darkening and any other chromatic effects that may be at play due to spots.

\begin{figure*}[th!]
\begin{center}
\includegraphics[width=0.98\textwidth]{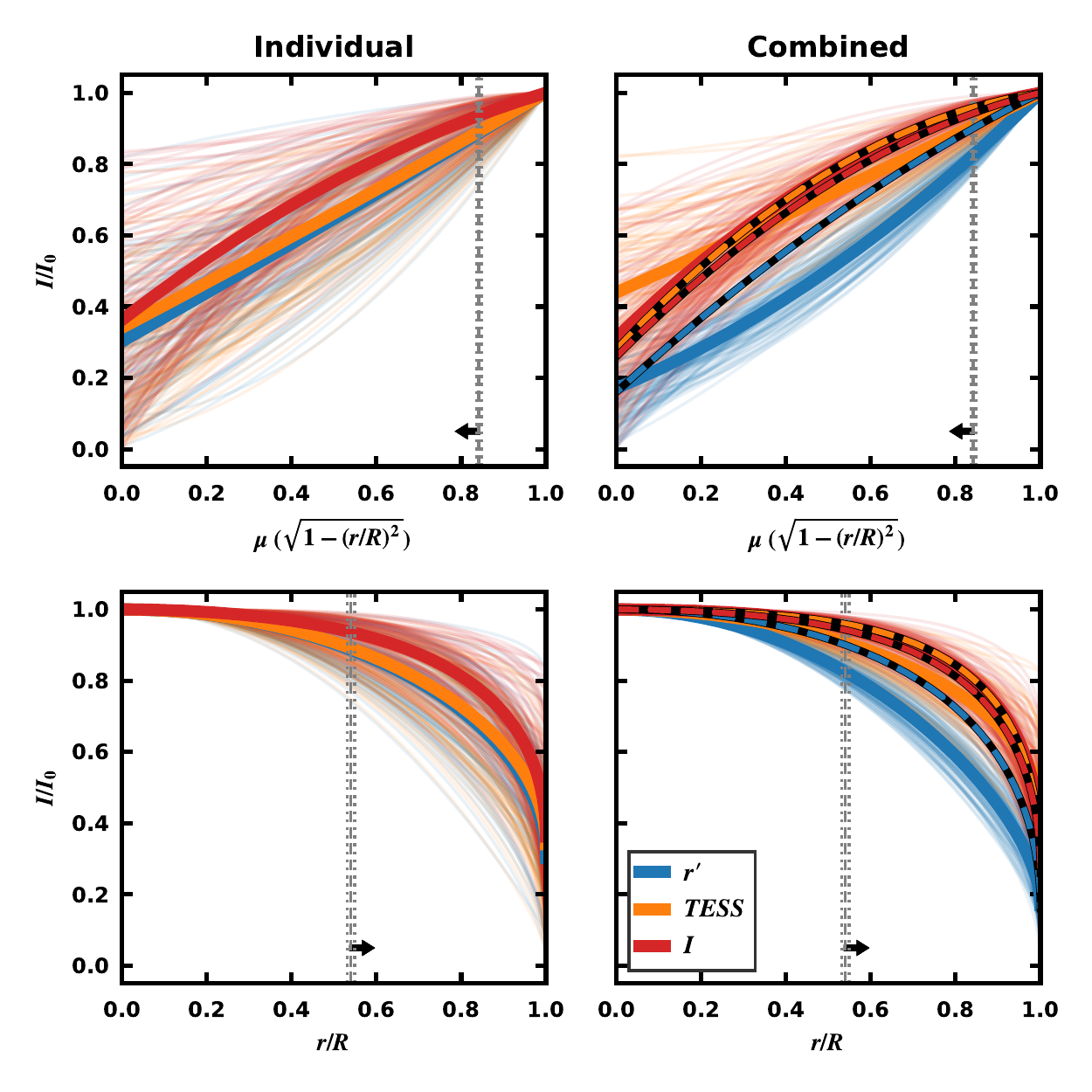}
\caption{Filter-dependent radial brightness profiles from our individual filter (left) and combined fits (right). Top panels present the surface brightness with respect to $\mu$. The bottom panels are with respect to the normalized radius coordinate. The median profile for each filter is shown with a thick line, and 50 random draws from the limb-darkening coefficient posteriors are shown in thin lines of the associated color. In the right panels, dashed lines with black backgrounds represent the theoretical predictions. The vertical gray lines mark the maximal radius occulted by the eclipse (solid) and its uncertainty (dashed), and arrows indicate the portion of the plot that correspond to eclipsed area.}
\label{fig:LD}
\end{center}
\end{figure*}

We find good agreement between the fit variations. The largest differences exist in the LDCs, whose values shift and become more constrained in the combined fit. The corresponding radial brightness profiles are presented in Figure \ref{fig:LD} and discussed further in Section \ref{LDCs}. The $r^\prime$ LDCs show the largest change between the individual and combined fit ($\sim$1$\sigma$). The difference has a negligible impact on the derived properties, in part, because the LDCs are poorly constrained in both fits. Most of the variation occurs between the orbital inclination ($i$) and normalized orbital separation ($a/(R_1+R_2)$), which are covariant while producing the same derived radii between the fits. 

In the light of the agreement between the fit variations, we adopt the combined fit as fiducial. The result is a stellar twin system consisting of two 0.177 $M_\odot$ stars with radii of 0.35 $R_\odot$ on a short-period (10.714762 d), eccentric orbit ($e = 0.2969$). Formal uncertainties on the masses are $\sim$0.2\%. Formal radii uncertainties are $\sim$1\%, but we address potential sources of systematic uncertainties in Section \ref{spots}. The radii are larger than the MS prediction, consistent with the our expectation that $\sim$40 Myr stars of this mass should still reside on the pre-MS. 

We note that the individual radii returned by our two-component, synthetic template fit in Section \ref{priors}, $0.358^{+0.008}_{-0.011}$ and $0.361^{+0.010}_{-0.008}$ \rsun, are systematically larger but have fair agreement (just over 1$\sigma$). From our empirical, single-component fit in Section \ref{sed}, we assume both stars have the same \teff\ and luminosity (reasonable given our results in this section). Using the \gaia\ distance to compute the bolometric luminosity, we compute radii of $0.348\pm0.023$ \rsun, in better agreement, albeit with a larger uncertainty.

\section{Assumptions in Eclipse Fitting}
\label{ecl_ass}

The largest assumptions made in our modeling of the binary eclipses are that the stellar surfaces are a single temperature (spot free) and that their radial brightness profiles can be described with a quadratic limb-darkening law. The former we know to be false given the rotational modulation seen in the \tess\ light curve (Figure \ref{fig:tess}) and the flux-ratio variability we observed in our SALT--HRS spectra (Figure \ref{fig:FR}). The latter may not be categorically false, but it has been shown that even if a star's radial brightness profile can be described by a quadratic law, the theoretical predictions have large systematic offsets for cool stars \citep{Patel&Espinoza2022}. In the following subsections, we attempt to determine the impact of these assumptions, particularly with respect to the derived stellar radii. 

\

\subsection{Limb Darkening}
\label{LDCs}

In Figure \ref{fig:LD} we present the best-fit, filter-dependent radial surface-brightness profiles for the individual filter fits (left) and the combined fit (right). The top panels present the profile with respect to $\mu$ ($\sqrt{1-(r/R_\star)^2}$); the bottom panels are plotted as a function of the normalized radius coordinate ($r/R_\star$). The vertical lines mark the innermost radius occulted during the eclipse, where our data are able to apply constraints. In the individual fits, we find that the LDCs are largely unconstrained, as shown by wide range of faint profiles, which are random draws from the LDC posteriors. Demanding that all filter light curves correspond to the same orbital and stellar parameters, as we do in the combined fit, we find the the LDCs are much more constrained and that the $r^\prime$ profile falls off more steeply. This difference affects the interplay between the orbital inclination and normalized orbital separation ($a/(R_1+R_2)$), but as discussed above (Section \ref{results}), they do not have a significant effect on the derived radii. The combined fit highlights the value of a simultaneous multicolor fit in determining accurate LDCs. 

In the right panels of Figure \ref{fig:LD} we also present the theoretical predictions for each filter in dashed black and colored lines. Values are the mean of predictions from \citet{Claret&Bloemen2011}, \citet{Claret2017}, and the Exoplanet Characterization Tool Kit \citep{Bourqueetal2021}. The $I$ band predictions are the only ones that agree with our fit values within 1$\sigma$; however, the $I$ and \tess\ curves generally trace the range of profiles allowed by our data. The largest discrepancy exists for $r^\prime$, which predicts a shallower fall off than our best-fit values. 

To determine the affect of simply assuming the theoretical values, or applying a narrow prior on the LDCs, as is sometimes done in transit and eclipse fitting, we perform a combined fit fixing the LDC to the predicted values ($u_{1,r^\prime} = 0.59$, $u_{2,r^\prime} = 0.25$, $u_{1,TESS} = 0.17$, $u_{2,TESS} = 0.55$, $u_{1,I} = 0.29$, $u_{2,I} = 0.45$). From this fit, the derived radii values ($R_1=0.352 \pm 0.003$, $R_2=0.348 \pm 0.005$) agree with the fiducial combined fit within the 68\% confidence intervals. The Bayesian information criteria (BIC) for these two models are equivalent (the fixed LDC model is 0.05\% lower), indicating that our data are just at the point where they are able to provide meaningful constraints on the LDCs. This may be because the grazing eclipse only probes a small fraction of the stellar radius, or because our photometry does not have the precision to capture the subtle variations in the eclipse shape due to limb-darkening. With these findings, we conclude that our assumption of a quadratic law, and whether the LDCs are adopted from theory or fit, do not affect our fiducial fit results. 

This discussion has not included the contribution from star spots, which is discussed in more detail in the following section, but an important caveat is worth including here. Briefly, because this system's eclipse is grazing, limb-darkening has a large effect on the the wavelength-dependent eclipse depth (shallower eclipses at longer wavelengths). The same behavior could be expected from occulting a heavily spotted area. Because these two effects are degenerate, and the spot orientation is unknown, the LDCs fit here should not be take as empirical truth for pre-MS M 4.5 stars, but rather the values that best account for the combined contribution of limb-darkening and this system's specific spot properties. This does not necessarily mean that the theoretical LDCs are correct; \citet{Patel&Espinoza2022} found systematic offsets even for inactive solar-type stars, but some of the larger offsets seen for low-mass stars may be inflated by the unaccounted presence of spots, which we discuss below.

\subsection{Star Spots}
\label{spots}

Star spots can alter the depths and detailed shapes of eclipse light curves. These effects are typically ignored but can produce biases in derived radii that are significantly larger than the typical $\sim$1\% formal uncertainties in an unspotted fit (e.g., Section \ref{results}). Star spot crossing events produce the most obvious effect by introducing structure into the eclipse light curve \citep[see][for examples of spotted EBs from \kepler]{Hanetal2019}. Less obvious and more problematic are the effects of uneclipsed spots, or the eclipse of large spot complexes, which can bias radius measurements \citep[e.g.,][]{Rackhametal2018}. Here we assume that spots are the dominant surface features, and that faculae and plages can be ignored. Young, active solar-type stars are found to be spot-dominated \citep{Montetetal2017}, which we assume extends to the active M stars in TOI 450.

The key parameter defining the direction and magnitude of the effect (deeper vs. shallower eclipses) is the ratio of the average, projected spot-covering fraction, \fspot, to the spot-covering fraction of the eclipsed area, \fspotecl. This ratio encodes relative flux that each region carries (eclipsed vs. uneclipsed), which determines the eclipse depth. For instance: 
\begin{enumerate}
    \item If the ratio is unity (\fspot\ $=$ \fspotecl), independent of the specific \fspot\ value, or the presence of discrete spot-crossing events, the average eclipse depth will be the same as an unspotted system.
    \item If the ratio is greater than one (\fspot\ $>$ \fspotecl), i.e., a less-spotted eclipsed area, the eclipse depth will increase compared to an unspotted model because the eclipsed region carries a larger relative share of the total flux.
    \item If the ratio is less than one (\fspot\ $<$ \fspotecl), i.e., a more-spotted eclipsed area, the eclipse depth will decrease compared to an unspotted model because the eclipsed region carries a smaller relative share of the total flux.
\end{enumerate}  
In transiting exoplanet systems, this is known as the transit light source effect \citep{Rackhametal2018,Rackhametal2019}, and has straightforward impacts on the derived planetary radii: transiting less-spotted areas bias radii to larger values; transiting more-spotted eclipse areas bias radii to smaller values. In EBs, predicting the effect that spots have on derived radii is less straightforward. Combinations of the radius ratio, surface-brightness ratio, inclination, and orbital separation can conspire to produce counterintuitive results that require detailed modeling. This further emphasizes the value of priors informed by spectroscopy to limit areas of parameter space (see Section \ref{priors}). Our ability to assess the impact of spots is also bolstered in this case with access to multicolor eclipse light curves. The change in the eclipse depth has a strong wavelength dependence, where any effect is more pronounced at shorter wavelengths where the spot contrast is larger. 

Measuring \fspot\ or \fspotecl\ is challenging in the best-case scenarios and is often not feasible. Light-curve variability amplitudes are only sensitive to the longitudinally asymmetric components of spots and generally underestimate the spot-covering fraction \citep{Rackhametal2018,Guoetal2018,Lugeretal2021}. Multicolor time-series photometry can diagnose the spot properties with wavelength-dependent modulation amplitudes, but with typical ground-based precision, this approach is only feasible for the most extreme spotted systems (T Tauri, RS CVn). NIR spectra can probe the projected spot-covering fraction through two-temperature spectral decomposition \citep[e.g.,][]{Gully-Santiagoetal2017,Gosnelletal2022,Caoetal2022}, but do not provide information on the spot orientation. Doppler imaging can map the distribution of hot and cold regions \citep{Vogtetal1999,Strassmeier2002}, but requires bright, rapidly rotating stars. Finally, NIR interferometry can reconstruct stellar surfaces, but it is limited to the closest stars with large angular sizes \citep{Roettenbacheretal2016}. All of these approaches are made more difficult in the presence of a binary companion. 

Without the data or means to constrain \fspot\ or \fspotecl\ directly, we begin by searching for temporal variability in the eclipse light curves caused by star spots. Visually, we do not find any coherent structures in the light-curve residuals and measure $\chi_{\rm red}^2$ values $\lesssim1$ for each of the eclipses (see Figure \ref{fig:eclipse}). The exception is the $I$-band eclipse ($\chi_{\rm red}^2 = 1.8$), which has deviations that are likely not astrophysical (e.g., variable cloud and/or water vapor opacity). They occur both in and out of eclipse and are not presented in the contemporaneous $r^\prime$ eclipse, where the signature of spots should be enhanced (shorter wavelength). For some of the individual \tess\ eclipses, the best fit appears systematically above or below the data (while still within the errors). This behavior could result from a variable spot-covering fraction between eclipses, which is plausible given the difference between the stellar rotation and orbital period (Figure \ref{fig:FR_epochs}). We perform a joint RV and eclipse light-curve fit for each individual \tess\ eclipse and compare the eclipse depth to our GP stellar variability model. Under simplified spot orientations, namely those where \fspot\ and \fspotecl\ are correlated with stellar rotation, the eclipse depth will correlate with the total flux. We do not find any significant trend between the two or any statistically significant variability in the \tess\ eclipse depth. From this analysis, at the precision of our data, we do not find evidence for spot-induced temporal variability in the eclipse events. 

To address how time-averaged spot properties may be biasing our derived radii, we perform additional fits to the combined data set ($r^\prime$, \tess, $I$, RVs) making various assumptions about the spot properties. In this approach, we scale the eclipse model by the ratio of the eclipse depth in a spotted scenario ($\delta_{\rm spot}$) to the eclipse depth without spots ($\delta_0$). Ignoring limb-darkening, which, to first order, will be same for a spot-free and spotted star, the spot-free primary eclipse depth is:
\begin{equation}
    \delta_0 = \frac{F_{\rm out}-F_{\rm ecl}}{F_{\rm out}} = \frac{\Omega_{\rm ecl}}{\Omega_2\left(\frac{J_2}{J_1}\right) + \Omega_1},
\end{equation}
where $F_{\rm out}$ and $F_{\rm ecl}$ are the fluxes out of eclipse and in eclipse, respectively. These are rewritten in terms of the projected surface area of the stars ($\Omega_1$, $\Omega_2$), the area of eclipsed region ($\Omega_{\rm ecl}$), and the stellar surface-brightness ratio ($J_2/J_1$). For a spotted system, the in- and out-of-eclipse fluxes now contain contributions from the spotted and ambient regions. In this case, the primary ellipse can be written as:
\begin{equation}
    \delta_{\rm spot} = \frac{\Omega_{\rm ecl,A}J_{\rm 1,A} + \Omega_{\rm ecl,S}J_{\rm 1,S}}{\Omega_{\rm 2,A}J_{\rm 2,A} + \Omega_{\rm 2,S}J_{\rm 2,S} + \Omega_{\rm 1,A}J_{\rm 1,A} + \Omega_{\rm 1,S}J_{\rm 1,S}}, 
\end{equation}
where the same notation holds, but is now subscripted by an ``S'' or ``A ''to indicate the spotted and ambient surfaces, respectively. To arrive at the desired quantity, we can divide these two equations resulting in: 
\begin{equation}
\frac{\delta_{\rm spot}}{\delta_0} = \frac{\left(1 + f_{\rm s, ecl}(C_1-1)\right) \left(\left(\frac{R_2}{R_1}\right)^2\left(\frac{J_2}{J_1}\right)+1\right)}{\left(1 + f_{\rm s,2}(C_2-1)\right)\left(\frac{R_2}{R_1}\right)^2\left(\frac{J_2}{J_1}\right) + 1 + f_{\rm s,1}(C_1-1)},
\label{eqn:tr_ratio}
\end{equation}
where we have simplified some variables to align with our eclipse fitting parameters. We replace $\Omega_2/\Omega_1$ with $(R_2/R_1)^2$, and define  $f_{\rm s, ecl}$ as the spot-covering fraction of the area eclipsed on the primary star ($\Omega_{\rm ecl,S}/\Omega_{\rm ecl}$), $C_1$ and $C_2$ as the ratio of the spotted to ambient surface brightness on the primary and secondary, respectively (e.g., $J_{\rm 1,S}/J_{\rm 1,A}$), and $f_{\rm s,1}$ and $f_{\rm s,2}$ as the spot-covering fractions of the primary and secondary, respectively (e.g., $\Omega_{\rm 1,S}/\Omega_{\rm 1}$). This ratio is filter specific as $J_2/J_1$, $C_1$, and $C_2$ are wavelength dependent. 

In its full form above, five additional fit parameters ($f_{\rm s, ecl}$, $f_{\rm s,1}$, $f_{\rm s,2}$, $C_1$, $C_2$) that scale the eclipse depth and that are largely degenerate with each other, are unlikely to be supported by present data. We can, however, make simplifying assumptions given our prior knowledge of the system that allow us to probe different extremes of the parameter space (Sections \ref{ecc_amb}, \ref{ecc_spots}), and allow for us to perform a fit of the spot properties under certain assumptions (Section \ref{fitspots}).

In each of the exercises, we leverage our knowledge of the TOI 450 stars and their similarity by pre-computing $C$, assuming it is the same for each star. We do so by combining model spectra from the {\tt BTSettl-CIFIST} suite \citep{Baraffeetal2015} and convolving them with each filter profile. We set an ambient photospheric temperature of 3100 K and a spot-photosphere temperature ratio of 0.92 \citep{Berdyugina2005,Afram&Berdyugina2015,Fangetal2018,Rackhametal2019}. For the $r^\prime$, \tess, and $I$ filters, we compute spot to ambient surface-brightness ratios of 0.29, 0.53, and 0.63, respectively.

\subsubsection{Eclipsing Ambient Photosphere}
\label{ecc_amb}
In this scenario we assume that the eclipse only passes over the ambient photosphere (\fspotecl $= 0$), but there exists some average spot filling factor. Here we assume that both stars have the same \fspot. Under these assumptions, Equation \ref{eqn:tr_ratio} can be simplified to:
\begin{equation}
\frac{\delta}{\delta_0} = \frac{1}{1+f_{\rm s}(C-1)}.
\label{eqn:unon_simp}
\end{equation}
Using Equation \ref{eqn:unon_simp} we select four spot-covering fractions (\fspot\ = 0.1, 0.2, 0.3, 0.4), scale the eclipse model by $\delta/\delta_0$, which is always $>1$, deepening the eclipse, and perform a combined fit. Table \ref{tab:spot_fits} presents a subset of the results of these fits for parameters of interest. Here we find that the derived radii increase with the spot-covering fraction while the inclination decreases (larger impact parameter) to maintain the same eclipse duration. At \fspot\ $=0.2$, the radii differ by more than 1$\sigma$. At \fspot\ $=0.4$ the radii have increased by more than 5\%. In all cases, the radius ratio is consistent with unity. Figure \ref{fig:spots} provides a graphical ``toy-model'' representation for each model at first contact, showing the corresponding eclipse light-curve model in the absence of spots and with spots. The radial brightness profiles for each model are provided in the bottom row. The comparison of these eclipse curves highlights the impact that spots have on eclipse depths and the variety of spot properties, orbital orientations, and derived radii that produce equivalent light curves. For the ``Eclipsing Ambient'' models specifically, we see that increasing \fspot\ deepens the eclipse (i.e., shallower in the ``without Spots'' row), and increases the difference between the eclipse depths in the different filters. The latter effect requires more exaggerated differences in the filter-dependent radial brightness profiles to match the observed eclipse depths. 

Despite their ability to reproduce the observe eclipse depths, there is circumstantial evidence to disfavor high \fspot\ values in this scenario. For instance, even in this grazing orientation, the primary eclipse covers roughly 12\% of the projected stellar surface, which makes high \fspot\ values contrived as to exclude spots from the eclipsed region. Also, higher \fspot\ values require increasingly steep radial bright profiles to match the observed eclipse depth. Even with significant systematic uncertainties in theoretical LDCs, the high \fspot\ radial brightness profiles are likely unphysical. 

Low \fspot\ models remain plausible. Including these possibilities requires roughly doubling the uncertainty in the derived radius from the fiducial fit. 

\subsubsection{Eclipsing Spots}
\label{ecc_spots}
In the opposite extreme, we assume that the all latitudes on the primary star below the highest extent of the grazing eclipse are spotted (i.e., \fspotecl\ $= 1$), while the rest of the primary and secondary are spot free ($f_{\rm s,2} = 0$). While this might represent a pathological spot orientation, polar spots are often observed in Doppler imaging studies \citep{Strassmeier2009}. The transit depth ratio for this case is:
\begin{equation}
\frac{\delta}{\delta_0} = \frac{\left(\left(\frac{R_2}{R_1}\right)^2\frac{J_2}{J_1} + 1\right)C}{\left(\frac{R_2}{R_1}\right)^2\frac{J_2}{J_1} + 1 - f_{\rm s} + f_{\rm s}C}.
\label{eqn:oc_simp}
\end{equation}
We perform a combined fit scaling the eclipse model by $\delta/\delta_0$, which is always $<1$, producing shallower eclipses for a given set of parameters. For this exercise, \fspot\ is dependent on the orbital and stellar parameters and is computed on-the-fly for each model. 

\begin{figure*}[th!]
\begin{center}
\includegraphics[width=0.98\textwidth]{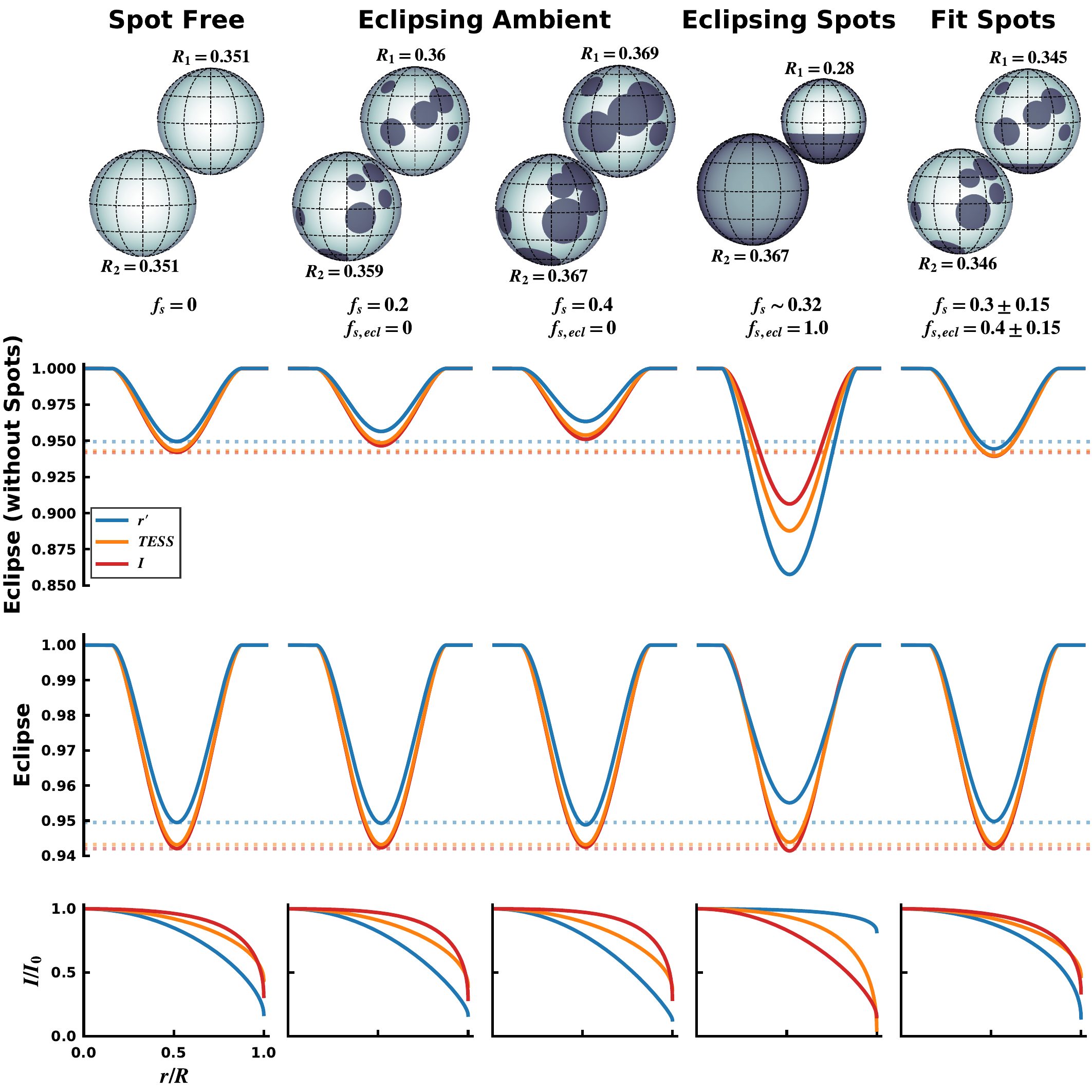}
\caption{Comparison of fits including spots. The top row presents a possible realization of the stellar surfaces at first contact of the primary eclipse, with the derived radii labeled for each star. The associated spot-covering fraction (\fspot) and spot-covering fraction of the eclipsed area ($f_{\rm s, \ ecl}$) are provided below. The second and third rows present the eclipse models without and with the effect of spots included for each filter, respectively. Horizontal dotted lines in these rows represent the best-fit eclipse depth from the fiducial fit (i.e., the observed eclipse depth; Figure \ref{fig:eclipse}). The last row presents the limb-darkening profiles for each filter.  The leftmost column is the fiducial fit (Section \ref{results}). The middle two columns are a subset of the ``eclipsing ambient'' models (Section \ref{ecc_amb}). The third column is the ``eclipsing spots'' model (Section \ref{ecc_spots}), which can be ruled out based on its poor match to the filter-dependent eclipse depths and its reversal of the expected filter-dependent limb-darkening trend. The secondary appears completely spotted, but instead has a lower surface brightness than the primary, which is favored in this fit. The rightmost column is the fit spot model (Section \ref{fitspots}), from which we adopt our definitive measurements. Diverse spot configurations and derived radii can reproduce the observed eclipse depths. } 
\label{fig:spots}
\end{center}
\end{figure*}

The large effect occulting a spotted region has on the eclipse depth sends the fit to extreme regions of the allowed radius-ratio and surface-brightness-ratio parameter space. Select parameters from the fit are presented in Table \ref{tab:spot_fits}, with a graphical representation in Figure \ref{fig:spots}. To balance the reduced eclipse depth, this fit reduces the surface-brightness ratio (less flux dilution from the secondary), and increases the relative occulted area by decreasing the primary radius by $\sim$20\% and decreasing the impact parameter (higher inclination; less grazing). The secondary appears fully covered in spots in Figure \ref{fig:spots}, but this is instead the realization of the extreme stellar surface-brightness ratio this fit prefers. The corresponding primary \fspot\ is $\sim0.32$. This fit resides in a much less likely area of the radius-ratio surface-brightness-ratio prior (Section \ref{priors}) compared to other fits above, but it is the multicolor eclipse information that allows us to completely rule out this scenario. Not only is the fit unable to reproduce the relative eclipse depths in $r^\prime$, \tess, and $I$ (Figure \ref{fig:spots}), the corresponding unspotted model predicts steeper limb-darkening at redder wavelengths, which is the opposite of theoretical predictions and empirical findings \citep[e.g.,][]{Mulleretal2013}.

\subsubsection{Fit Spots}
\label{fitspots}
In this last scenario, we assume the spot-covering fraction of both stars are the same ($f_{\rm s} = f_{\rm s,1} = f_{\rm s,2}$) and allow both \fspot\ and \fspotecl\ to be fit as free parameters. With this setup, the eclipse depth ratio becomes, 
\begin{equation}
\frac{\delta}{\delta_0} = \frac{1 + f_{\rm s,ecl}(C-1)}{1+f_{\rm s}(C-1)},
\label{eqn:fit_simp}
\end{equation}
which we use to scale the eclipse depth. Although \fspot\ is unknown for TOI 450, we place a prior on its value based on the spot-covering fractions measured from SDSS-APOGEE 

\

\

\noindent spectra \citep{Majewskietal2017} of young cluster members \citep[][Cao, L.\ private communication]{Cottaaretal2014,Donoretal2018,Caoetal2022}. The decreasing trend with age predicts \fspot\ $\sim 0.3$ for an age of 40 Myr, which we adopt as the center of our normal distribution prior with a width of 0.15 ($\mathcal{N}(0.30,0.15)$), allowing support for \fspot\ $=0$ models. In addition to this prior, \fspot\ and \fspotecl\ are limited to values between zero and one (\fspotecl\ does not have an informed prior). This approach is similar to that developed by \citet{Irwinetal2011} in its effect on eclipse depths, but it does not attempt to match the out-of-eclipse variability or absolute flux values between filters as we are working with normalized fluxes. 

We perform a combined fit for this scenario and present its results in Table \ref{tab:spot_fits} and Figure \ref{fig:spots}. The eclipses themselves do not constrain \fspot, and as such, our fit returns the input prior. The fit does constrain \fspotecl, however, returning a value of $0.39\pm0.15$. The spot parameter posteriors are positively correlated, and correspond to an \fspotecl$/$\fspot\ value of $1.4^{+0.7}_{-1.1}$. The change in the eclipse depth in the \tess\ bandpass corresponds to  $0.94 \pm 0.05$. The stellar radii for this model are less than the fiducial, but still consistent, owing to the larger uncertainty in this spotted model ($\sim2$\% precision). The BIC for this model is marginally higher than the fiducial fit (0.02\%), but not significantly different as to rule out its use.  

In this scenario, the fit favors models in which spots act to shallow the eclipse depth and reduce the difference in eclipse depth across the three filters. The LDCs of this fit are in better agreement with the theoretical predictions, where both the $r^\prime$ and $I$ values are in agreement, and while the \tess\ values are not in strict agreement, they generally trace the same radial brightness profile. This result may be signifying that the sharp radial brightness profiles required to match the eclipse depths in the absence of spots provide a worse match to the eclipse shape, and reducing the eclipse depth with spots allows the LDCs to more easily describe the eclipse shape. This distinction is largely possible because we are able to jointly constrain the LDCs of three filters simultaneously. 

To test the impact of the assumed \fspot\ prior, we perform an additional fit with a narrower and lower \fspot\ prior, $\mathcal{N}(0.1,0.1)$. This fit returns consistent stellar and orbital parameters with the previous fit. As before, the \fspot\ posterior returns the prior. The \fspotecl\ value is higher in this fit, $0.26 \pm 0.13$, but the \fspot\ and \fspotecl\ pair result in the same transit depth ratio. This exercise reveals that our approach does not constrain the spot properties themselves, only whether the fit favors an eclipsed area that is more or less spotted than the global average. 

We perform two additional tests to assess the impact of our choice of the limb-darkening prescription and the spot-to-ambient temperature contrast. In the first, we implement a square-root limb-darkening law \citep{Klinglesmithetal1970}, which has been show to provide a better approximation of the NIR stellar intensity profile of late-type stars \citep{vanHamme1993}. We do not select this limb-darkening law in the fits above because it does not have an analytical implementation in {\tt batman} and is too computationally expensive for the variety of fits we have explored. With the $\mathcal{N}(0.30,0.15)$ \fspot\ prior, we derive radii of $0.344^{+0.004}_{-0.005}$ and $0.343^{+0.006}_{-0.005}$ for the primary and secondary, respectively, in good agreement with the quadratic limb-darkening result above (Table \ref{tab:spot_fits}). Lastly, we perform two quadratic limb-darkening fits ($P(f_{\rm s}) = \mathcal{N}(0.30,0.15)$), setting the spot-to-ambient temperature contrast to 0.89 and 0.95, as opposed to 0.92 used above. These result in the following: $R_1 = 0.344 \pm 0.007$ and $R_2 = 0.345^{+0.007}_{-0.006}$ for $T_{\rm spot}/T_{\rm amb}$ = 0.89 and $R_1 = 0.347^{+0.006}_{-0.005}$ and $R_2 = 0.347^{+0.006}_{-0.005}$ for $T_{\rm spot}/T_{\rm amb}$ = 0.95. In each of these fit variations, the derived radii are consistent with the initial fit in this Section within 1$\sigma$.

\begin{deluxetable*}{l c c c c c c c}
\tablecaption{Comparison of Spotted Models
\label{tab:spot_fits}}
\tablewidth{0pt}
\tabletypesize{\footnotesize}
\tablecolumns{8}
\phd
\tablehead{
  \colhead{Parameter} &
  \colhead{No Spots} &
  \multicolumn{4}{c}{Eclipsing Ambient} &
  \colhead{Eclipsing Spots$^b$} &
  \colhead{Fit Spots$^c$}\\
  \colhead{} &
  \colhead{(Fiducial Fit)} &
  \colhead{} &
  \colhead{} &
  \colhead{} &
  \colhead{} &
  \colhead{} &
  \colhead{(Definitive Fit)}\\
  \cline{3-6}
}
\startdata
\fspot$_{,1}$     & 0.0                       & 0.1                       & 0.2               & 0.3$^a$           & 0.4$^a$                  & $\sim0.32$                & 0.30$\pm$0.15     \\
\fspot$_{,2}$     & 0.0                       & 0.1                       & 0.2               & 0.3               & 0.4                      & 0.0                       & 0.30$\pm$0.15     \\
\fspotecl         & 0.0                       & 0.0                       & 0.0               & 0.0               & 0.0                      & 1.0                       & 0.39$\pm$0.15     \\
\hline
\\
$R_1 (R_\odot)$   & 0.351$\pm$0.003           & 0.355$\pm$0.003           & 0.360$\pm$0.003   & 0.364$\pm$0.003   & $0.369^{+0.004}_{-0.003}$& $0.280^{+0.004}_{-0.003}$ & 0.345$\pm$0.006   \\
$R_2 (R_\odot)$   & $0.351^{+0.005}_{-0.004}$ & $0.355^{+0.005}_{-0.004}$ & 0.359$\pm$0.005   & 0.363$\pm$0.005   & 0.367$\pm$0.005          & $0.367^{+0.004}_{-0.003}$    & 0.346$\pm$0.006   \\
$R_2/R_1$         & 1.00$\pm$0.02             & 1.00$\pm$0.02             & 1.00$\pm$0.02     & 1.00$\pm$0.02     & 1.00$\pm$0.02            & $1.311^{+0.007}_{-0.008}$ & 1.00$\pm$0.02     \\
$J_2/J_1 (\tess)$ & 1.00$\pm$0.03             & $1.00^{+0.03}_{-0.02}$    & 1.00$\pm$0.03     & 1.01$\pm$0.03     & 1.01$\pm$0.03            & $0.673^{+0.010}_{-0.008}$ & 1.01$\pm$0.03     \\
$i$ (degrees)     & 87.29$\pm$0.02            & 87.24$\pm$0.02            & 87.19$\pm$0.02    & 87.13$\pm$0.02    & 87.08$\pm$0.02           & 87.74$\pm$0.02            & 87.35$\pm$0.06    \\
$M_1 (M_\odot)$   & 0.1768$\pm$0.0004         & 0.1768$\pm$0.0004         & 0.1768$\pm$0.0004 & 0.1768$\pm$0.0004 & 0.1769$\pm$0.0004        & 0.1765$\pm$0.0004         & 0.1768$\pm$0.0004 \\
$M_2 (M_\odot)$   & 0.1767$\pm$0.0003         & 0.1768$\pm$0.0003         & 0.1768$\pm$0.0003 & 0.1768$\pm$0.0003 & 0.1768$\pm$0.0003        & 0.1765$\pm$0.0003         & 0.1767$\pm$0.0003 \\
\enddata
\tablenotetext{a}{Models disfavored based on their stellar radial brightness profiles and contrived spot geometries.}
\tablenotetext{b}{Model completely ruled out by multi-wavelength eclipse light curves.}
\tablenotetext{c}{Adopted definitive fit.}
\end{deluxetable*}

\subsection{Adopted Stellar Radii \& Spot Summary}
\label{adpoted}

We adopt the results of the spotted fit in Section \ref{fitspots} (\fspot $=\mathcal{N}(0.30,0.15)$) as our definitive measurement, which returns radii of 0.345 and 0.346 $R_\odot$ for the primary and secondary, respectively, with a formal uncertainty of 0.006 $R_\odot$ ($\sim$2\% precision). These values are robust to our choice of limb-darkening profile, and spot properties. (Other fit and derived values that differ significantly from the fiducial fit are included in Table \ref{tab:spot_fits}.) We note that stellar masses are independent of any plausible spot model explored, owing to its weak $\sin^3i$ dependence at high inclinations. This approach includes the effect of spots under minimal added model complexity (two additional parameters) and modest assumptions: spots exist on the stellar surfaces, the spot properties of the primary and secondary are the same, and the spot-covering fraction of the eclipsed area can be different than the average, projected value. Regardless of the specific \fspot\ value, the fact that this grazing eclipse favors \fspotecl\ $>$ \fspot\ values, points to a distribution of spots that favors high latitudes (i.e., more polar than equatorial configurations). 

Throughout Section \ref{spots} we have shown that spots can produce significant changes in derived radii. The effects spots have on eclipse depth are largely degenerate with limb-darkening. Allowing the LDCs to vary can mask the effects of spots, producing a wide variety of spot orientations that are consistent with observations. Multicolor eclipse light curves provide important additional constraints that can narrow the range of allowed spot properties. Our analysis finds that including a flexible spot prescription results in best-fit LDC values that are in good agreement with theoretical predictions. This result suggests that rigorous tests of the limb-darkening models require multicolor observations that include the effect of spots. 

In the case of TOI 450, our spot prescription results in a reduction of the stellar radius on the order of $\sim$2\% from the unspotted, fiducial model. This would seem to ease the tension between model radii and observations. However, it should not be assumed that the result here will apply to all EB systems. The direction that spots may be biasing derived radii will be unique to each system. TOI 450's grazing orientation likely makes it more susceptible to this effect. Systems with lower impact parameters, where the area eclipsed is a larger fraction of the total projected area, are less likely to result in \fspotecl\ values that differ significantly from global spot-covering fraction. While including the effect of spots is important for obtaining accurate radii and realistic uncertainties, we do not suggest that the tension between observed and model radii can be resolved with spots. This finding is also supported by the spotted EB analysis in \citet{Irwinetal2011}.

\begin{figure*}[th!]
\begin{center}
\includegraphics[width=0.98\textwidth]{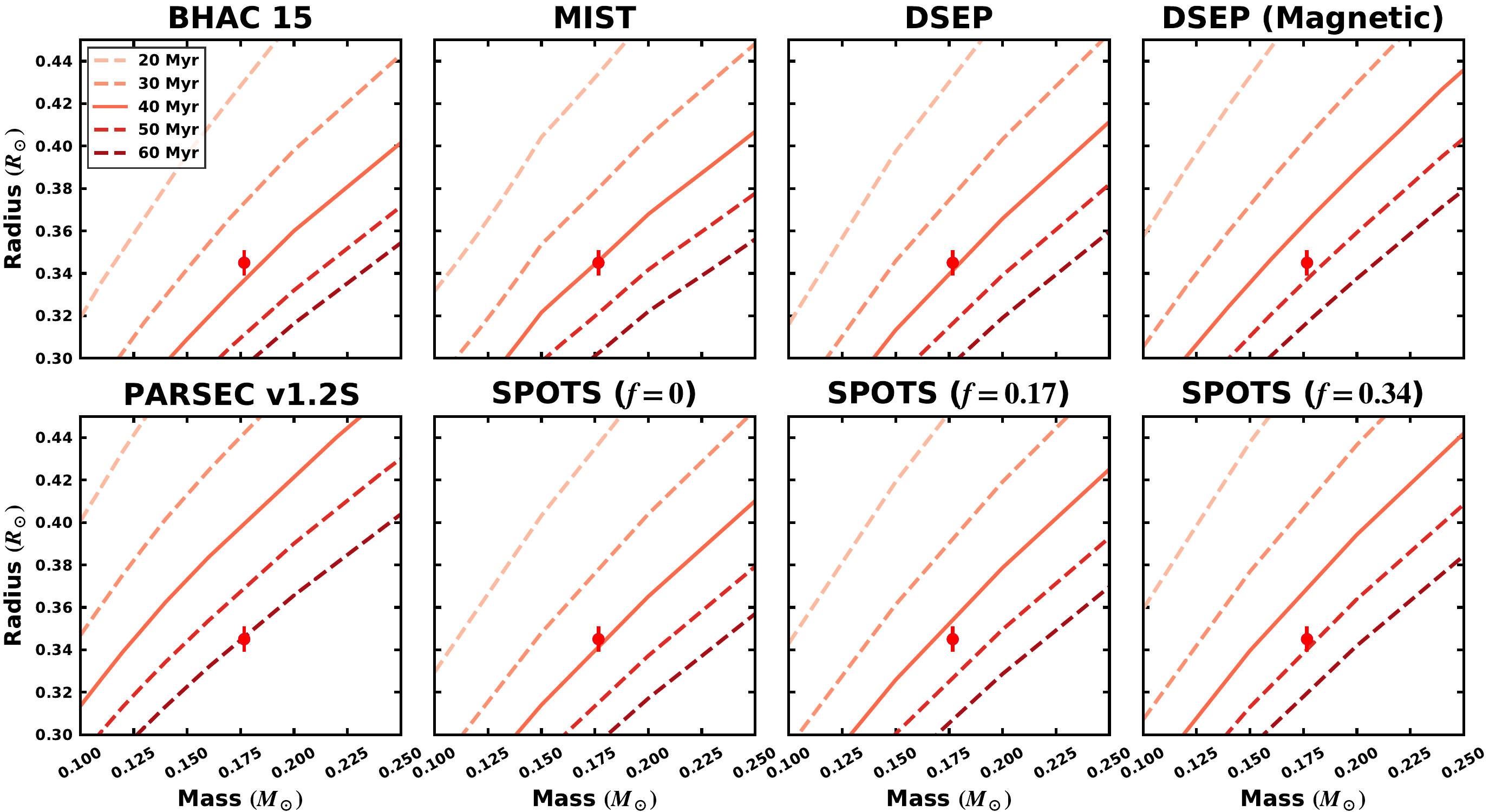}
\caption{Comparison of the TOI 450 masses and radii to isochrones from various stellar evolution model suites. See the text for model descriptions. Colored lines represent isochrones at specific ages provided in the legend. The literature age of Columba (40 Myr) is shown with a solid line. The TOI 450 primary and secondary are overlapping.}
\label{fig:MR}
\end{center}
\end{figure*}

\section{Comparison to Stellar Evolution Models}
\label{models}

To further constrain the age of the Columba association and to test models of pre-MS evolution, we compare our measurements of TOI 450 to various model isochrones. We select three standard stellar evolution models: BHAC 2015 \citep{Baraffeetal2015},  the Dartmouth Stellar Evolution Program \citep[DSEP,][]{Dotteretal2008}, and the MESA Iscochrones and Stellar Tracks \citep[MIST,][]{Dotteretal2016,Choietal2016}. We also select three additional model suites that attempt to correct for the shortcomings of standard models. The PAdova and TRieste Stellar Evolution Code \citep[PARSEC,][]{Bressanetal2012} version 1.2S \citep{Chenetal2014} introduces an ad hoc relation between the \teff\ and Rosseland mean optical depth to improve the agreement with mass-radius relation for dwarf stars. The DSEP magnetic models \citep{Feiden&Chaboyer2012,Feiden&Chaboyer2013,Feiden2016} include a prescription for magnetically inhibited convection which slows pre-MS contraction. We test the version that applies a magnetic field strength in equipartition with the thermal energy. And lastly, the Stellar Parameters Of Tracks with Starspots models \citep[SPOTS,][]{Somersetal2020} includes a star spot prescription that impedes the energy transport near the surface, inflating the stellar radii. We explore the \fspot\ $=$ 0, 0.17, and 0.34 versions. In all models, we assume solar metallicity. 

Figure \ref{fig:MR} presents the mass-radius (MR) diagram comparing the TOI 450 components to the models described above. With the exception of PARSEC v1.2S, all of the models predict ages between 30 and 50 Myr, in good agreement with the our expectation for a Columba member. The standard models (BHAC 15, MIST, DSEP, SPOTS ($f_s = 0$)) predict ages at or slightly below 40 Myr, while the DSEP Magnetic and spotted SPOT models ($f_s = 0.17$, $f_s = 0.34$) suggest older ages, between 40 and 50 Myr. The poor performance of the PARSEC v1.2s models may be the result of model alterations that are tailored to improve CMD agreement at field ages that do not carry over to young ages in MR space. 

\begin{figure*}[th!]
\begin{center}
\includegraphics[width=0.98\textwidth]{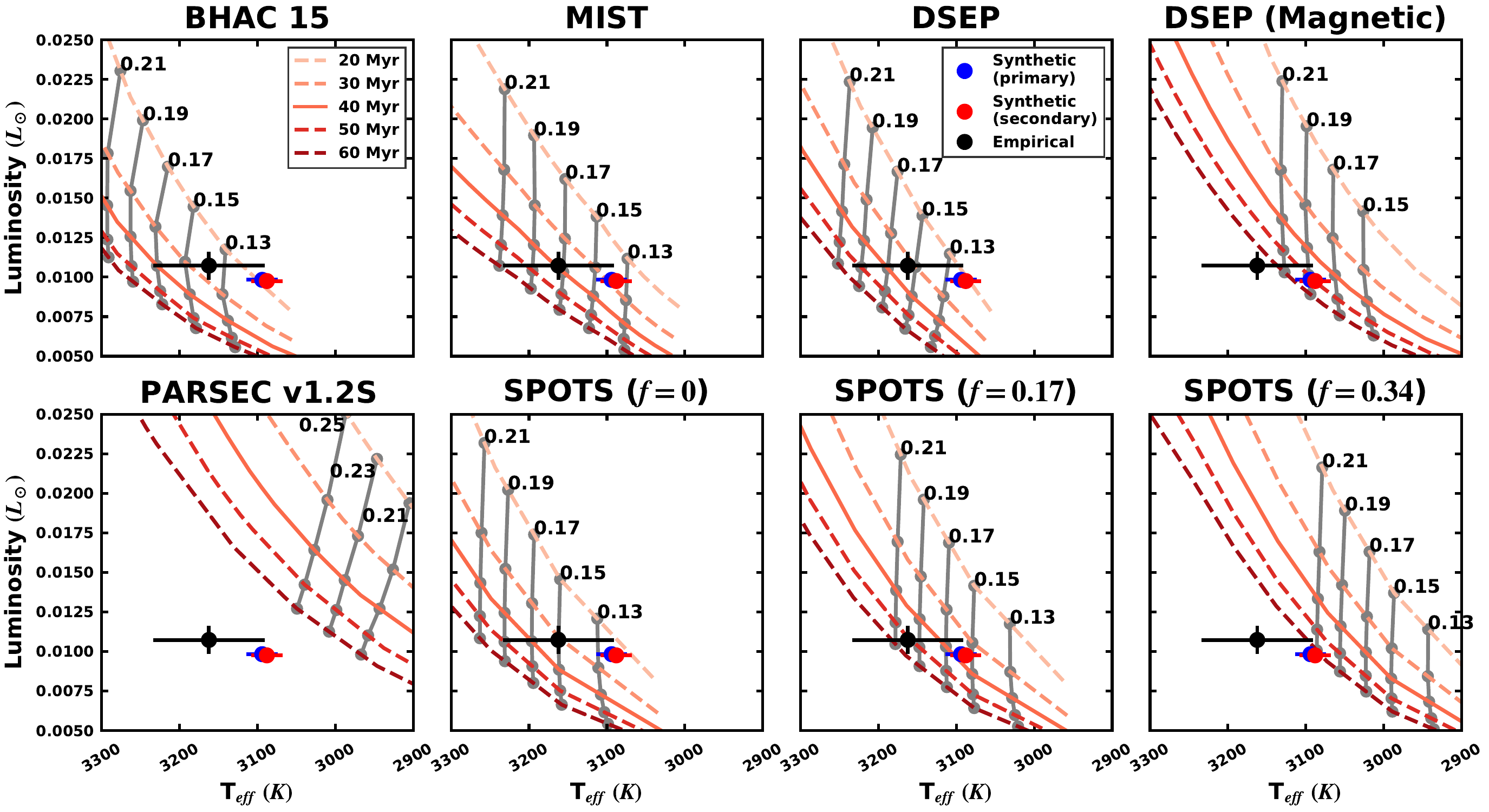}
\caption{Comparison of TOI 450 to isochrones from various stellar evolution model suits in the HR Diagram. See text for model descriptions. Colored lines represent isochrones at specific ages provided in the legend. The literature age of Columba (40 Myr) is shown with a solid line. Gray lines mark mass sequences, which are labeled in solar masses at the 20 Myr isochrone. The blue and red points are the \teff\ and luminosity for the primary and secondary, respectively, from the fit of synthetic templates described in Section \ref{priors}. The black point is a fit using empirical stellar templates (Section \ref{sed}), where we assume both stars have the same \teff\ and radius.}
\label{fig:HRD}
\end{center}
\end{figure*}

In Figure \ref{fig:HRD} we make the same comparison, now in the Hertzsprung-Russell (HR; \teff\ -- luminosity) diagram. We compute the \teff\ and luminosity using two approaches. In the first, we adopt the bolometric flux from the empirical-template SED fit in Section \ref{sed} and compute the bolometric luminosity assuming the \gaia\ distance. We then assume both stars have the same luminosity (i.e., divide by two) and compute the \teff\ using the Stefan–Boltzmann law and the derived radii from our EB fit (Section \ref{adpoted}). This results in the single black point in Figure \ref{fig:HRD}, whose error bar encompasses the positions of both stars. In the second approach, we adopt the \teff\ values from the two-component fit of synthetic model spectra to the SED and spectroscopic flux ratios (Section \ref{priors}). The bolometric luminosity is then computed via the Stefan–Boltzmann law using the radii from the EB fit. The formal uncertainties are small enough that we include the measurements for the primary and secondary separately as the blue and red points, respectively. We favor the former, {\it empirical} approach as it is less model dependent, but include both, as the latter, {\it synthetic} approach is common in the literature \citep[e.g.,][]{Davidetal2019}.

The HR diagram results follow the same trends seen in the MR diagram, but with a larger spread. Standard models (BHAC 15, MIST, DSEP, SPOTS ($f_s = 0$)) predict ages from 20 to 40 Myr. The empirical measurement approach (black circles) provide better agreement with the mass tracks shown with gray lines. The SPOTS ($f_s = 0.17$, $f_s = 0.34$) and DSEP Magnetic models span ages of 40 to 90 Myr and produce better mass agreement with the cooler, synthetic measurement approach (red and blue circles). PARSEC v1.2s models produce large offsets in this plane as well, predicting ages $>$100 Myr and masses $>$0.25 $M_\odot$.

To make a quantitative comparison with the models, we perform a two- or one-dimensional interpolation of the models depending on the comparison in question. To determine the model ages and masses, we use {\tt scipy.interpolate.gridddata} to linearly interpolate the MR and HR diagram model planes. For model radii and \teff\ for our measured dynamical mass, we use {\tt scipy}'s {\tt interp1d} for the one-dimensional interpolation. In each case, we pass normal distributions to the functions representing each measurement's value and uncertainty, taking the returned distribution's median and standard deviation as the model value and error.

\begin{figure*}[h!]
\includegraphics[width=0.98\textwidth]{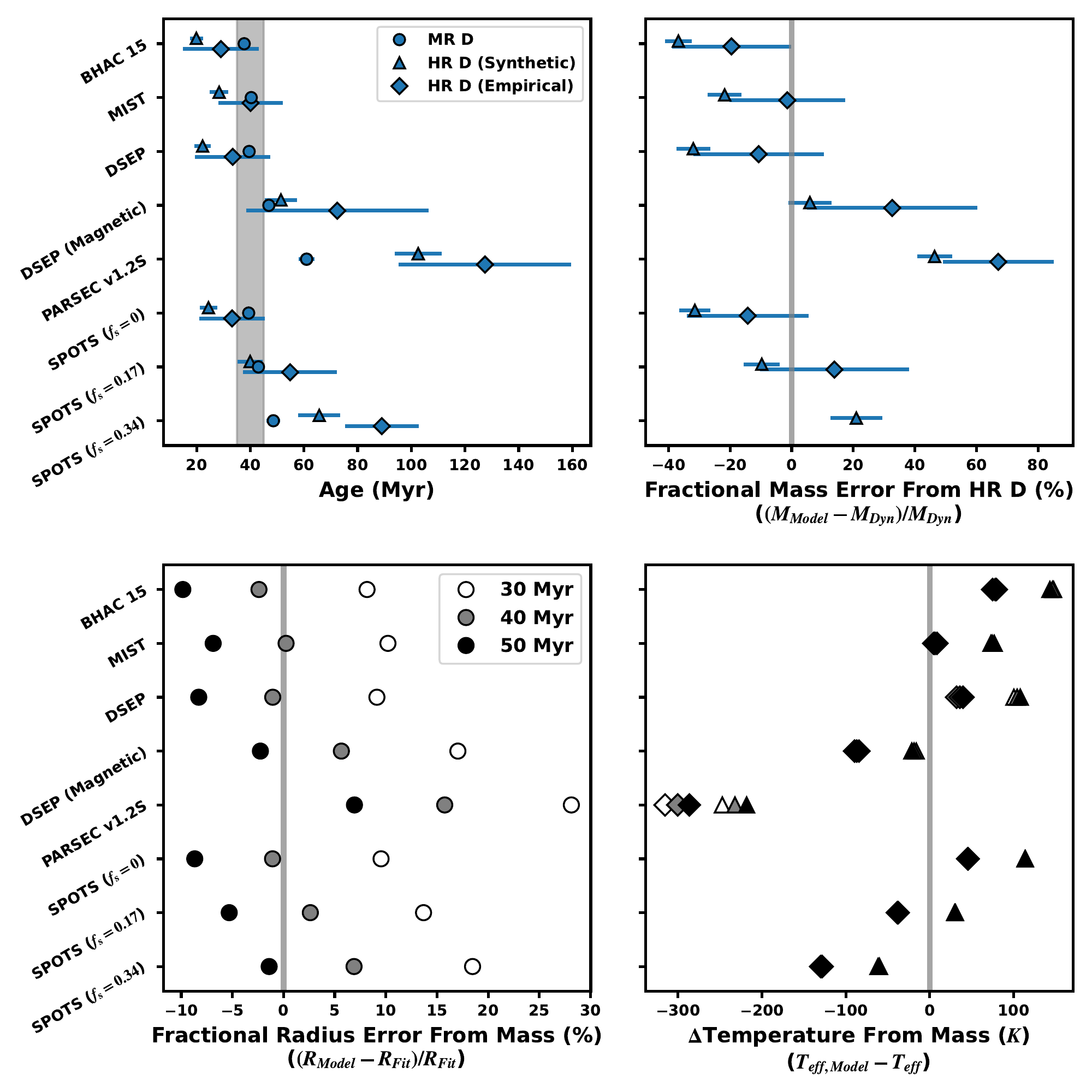}
\caption{Quantitative comparison of TOI 450 measurements with various model suites. In each panel, model suites correspond to rows of measurements. Measurements that are made in the mass-radius plane are depicted as circles. Measurements from the HR diagram are shown as diamonds and triangles representing empirical and synthetic measurement approaches, respectively. {\bf Top Left:} Isochronal age measurements. The literature Columba age, $\sim$40 Myr, is shown as the vertical shaded region. Measurements from different approaches are slightly offset vertically for clarity. {\bf Top Right:} Fractional difference between the HR diagram mass measurement and the dynamical mass. {\bf Bottom Left:} Fractional difference between the model radii given the dynamical mass and the derived radii from the EB fit. The comparison is made at three discrete ages. {\bf Bottom Right:} Absolute difference in \teff\ given the dynamical mass. Comparison is made at three discrete ages for each \teff\ measurement approach (empirical, synthetic). }
\label{fig:ages}
\end{figure*}

In Figure \ref{fig:ages} we present this quantitative comparison for parameters of interest. The top-left panel presents the model ages for different approaches (i.e., MR diagram, HR diagram empirical and synthetic) compared to the Columba literature age ($\sim$40 Myr). As discussed above, the MR diagram age values are largely consistent with each other and a Columba age, while the HR diagram values show a larger scatter, still centered around $\sim 40$ Myr. The top-right panel presents the model mass based on the HR diagram, finding our empirical measurement approach performs the best across different models with typical fractional uncertainties of $\sim$20\%. 

In the bottom two panels, we leverage the high precision of our dynamical mass measurement to test the accuracy of models in predicting radii and effective temperatures. The bottom-left panel displays the fractional radius difference for various models at three discrete ages. At 40 Myr, most models have good agreement, predicting the radius to within $\pm$5\%. The DSEP Magnetic and \fspot\ $= 0.34$ SPOT models predict larger radii than we measure, $>$5\%, but would provide better agreement if the Columba age were closer to 50 Myr. In the bottom-right panel, we present the absolute \teff\ difference for the dynamical mass, again at three discrete ages. The standard models predict \teff\ values slightly hotter than we observe ($\lesssim$100 K), while spotted and magnetic models predict cooler temperatures, still within $\sim$100 K agreement. 

Despite the high-precision measurements we have obtained for TOI 450, our test of stellar evolution models is hampered by the lack of a precise, independent age and the narrow area of parameter space we are probing with a twin system. This fact is made clear by the comprehensive analysis of coeval Upper Sco EBs by \citet{Davidetal2019}, whose components span masses from 0.1 to 5 \msun\ at 5-7 Myr. Their analysis highlights that model agreement is mass dependent, where most models perform best at intermediate masses ($0.3 M_\odot < M < 1 M_\odot$) and begin to diverge at lower masses. Critically, many of the models diverge in the same way, namely predicting older ages, or equivalently {\it larger} radii at a given age for low-mass stars (the opposite direction of the standard radius inflation problem). It is unknown whether this behavior continues at $\sim$40 Myr, so comparing models at TOI 450's masses (0.18 \msun) alone does not probe the existence of systematic offsets that may be overpredicting the ages of stars at this mass. It is worth noting that \citet{Davidetal2019} found the SPOTS models \citep[in their initial implementation;][]{Somers&Pinsonneault2015} produce the most consistent ages across the mass range explored. The fact that their values do not differ greatly from other models we compare to may signal that the effect is smaller or absent at $\sim$40 Myr. A precise Columba age (e.g., Li depletion boundary measurement) or additional EBs coeval with TOI 450 will be required to test this behavior. 

Lastly, we note that model agreement depends on the quantities being compared. All models perform the best in the MR plane and diverge in other comparisons that rely on detailed radiative transfer physics that differ between codes. For instance, the MIST models appear to perform better that the DSEP models in our tests, but \citet{Mannetal2019} founds the opposite in a mass-luminosity ($M_K$) comparison for field M dwarfs. Additionally, the PARSEC v1.2S and DSEP Magnetic models appear to perform equally well in a CMD comparison of the $\sim$11 Myr Musca association, as well as other intermediate-age ($\sim$200 Myr) populations \citep{Mannetal2022}, but diverge significantly here. 

In this light, our analysis emphasizes that pre-MS stellar evolution hosts additional challenges beyond the standard, MS radius inflation problem, particularly at low masses.

\section{Conclusions}

In this work we have characterized TOI 450, a young eclipsing binary in the $\sim$40 Myr Columba association. Our analysis makes use of multicolor eclipse light curves, allowing us to include the effect of star spots in our eclipse modeling, producing accurate stellar radii with realistic uncertainties. We compare our results to various stellar evolution models to assess their accuracy and refine the age of the Columba association. The conclusions of our study are as follows:

\begin{enumerate}
    \item {\it TOI 450 is a low-mass eclipsing binary.} From our followup observations of the nominal exoplanet candidate, consisting of high-angular-resolution imaging and time-series, high-resolution spectroscopy, we find that TOI 450 is an eccentric, near-equal-mass binary whose on-sky orientation produces only a single grazing eclipse. We do not find evidence for additional stellar sources in the system, bound or otherwise. The stars have M4.5 spectral types and effective temperatures of $\sim$3100 K.
    
    \item {\it TOI 450 is a member of the $\sim$40 Myr Columba association.} We confirm the BANYAN $\Sigma$ membership of TOI 450 to the Columba association using the {\tt FriendFinder}. This tailored search for coeval, comoving, companions is motivated Columba's diffuse on-sky clustering, which can lead to high contamination. Our search recovers many bona fide members of the Columba association, whose CMD distribution is consistent with the 40 Myr Tuc-Hor sequence. 
     
    \item {\it Priors from high-resolution spectra enable a precise eclipsing binary fit, despite its single eclipse.} Wavelength-dependent flux ratios across the orders of our SALT-HRS spectra, combined with the unresolved SED, are fit with a two-component model to jointly constrain the surface-brightness and radius ratio for the system. This fit is used to construct a joint surface-brightness-ratio--radius-ratio prior using a Gaussian kernel-density estimate. These parameters would normally be adequately constrained by a primary and secondary eclipse. With this approach, we achieve measurement precisions in this single-eclipsing system that are on par with double-eclipsing systems.
    
    \item {\it TOI 450 is a twin system on the pre-MS.} From our fiducial fit to the system, both stars are indistinguishable at our precision. We derive masses of 0.177 \msun\ with radii of 0.35 \rsun, placing these stars well above the MS expectation. 
    
    \item {\it Including the effect of star spots in our eclipse model results in a 2\% reduction in the stellar radii.} We include a parameterization of the effect of spots in our eclipse model that functionally acts to scale the eclipse depth. The direction and magnitude of this scaling depends on whether the spot-covering fraction of the eclipsed area is higher or lower than the global value, resulting in a shallower or deeper eclipse, respectively. Degeneracies between star spots and the stellar limb-darkening profile are reduced with multicolor eclipse light curves. We find that the eclipses favor a model in which the grazing eclipse occults a more heavily spotted area than the average, projected value. Without constraining the total spot-covering fraction, this result suggests that spots on the primary star are preferentially at high (absolute) latitudes. The derived radii are below the fiducial value by more than 2$\sigma$. This result is not representative for spotted EBs generally, and is not a solution to the so-called radius inflation problem. 
    
    \item {\it Model Comparisons.} Standard stellar evolution models (BHAC 15, MIST, DSEP, SPOTS (\fspot = 0)) perform well in describing the properties of TOI 450, assuming an age of 40 Myr. Masses measured from the HR diagram are systematically low but within error of our empirically derived luminosity and \teff\ values. Predicted radii at our dynamical mass are consistent within 5\% and predicted \teff\ values agree within 100 K. The \fspot\ = 0.17 SPOTS model performs equally well. Higher \fspot\ SPOTS models, the DSEP Magnetic models, and especially the PARSEC v1.2S models predict older ages, higher masses, and cooler temperatures than we observe, and are generally disfavored by our measurements. For this stellar mass and age, we find the MIST and SPOTS \fspot $= 0.17$ models provide the most consistent results across the tests we perform. We note that this is result is only valid for this mass and age and is not necessarily expected to extend to other mass and age regimes, or to agreement in the CMD.
\end{enumerate}

In this study we lay out a flexible framework for including the effect of spots when modeling EB eclipse light curves. The method benefits significantly from multicolor eclipse light curves that help to break degeneracies with limb-darkening. Our approach is complementary to others addressing the effect of spots (e.g., \citealt{Windmilleretal2010}, {\tt eb} \citealt{Irwinetal2011}; {\tt starry} \citealt{Lugeretal2019}) but does not require long baseline light curves or the assumption that the detailed spot pattern is unchanging. By allowing for various spot orientations in this modeling, we probe the potential for systematic offsets in derived radii and produce more conservative formal uncertainties that should ease the tension that exists between different groups modeling the same systems. Our analysis suggests spots introduce a $\sim2\%$ precision floor in derived radii when multicolor eclipse light curves are available, and are likely higher when fitting spotted systems with a single band. We hope this approach will provide more robust empirical measurements to test models, but ultimately, a larger population of benchmark EBs across age and mass is required to identify specific shortcomings and improve the next generation of stellar evolution models. 

\acknowledgments

BMT would like to thank Ronan Kerr for discussions of the Columba stellar population, Lyra Cao for discussion of the spot-covering fractions, the Heising-Simons Foundation, TESS Cycle 3 GI grant \#80NSSC21K0780, TESS Cycle 4 GI grant \#80NSSC22K0302, and NSF grant \#1716495 (PI ALK).

Funding for the TESS mission is provided by NASA's Science Mission directorate. We acknowledge the use of public TESS Alert data from pipelines at the TESS Science Office and at the TESS Science Processing Operations Center. This paper includes data collected by the TESS mission, which are publicly available from the Mikulski Archive for Space Telescopes (MAST). Resources supporting this work were provided by the NASA High-End Computing (HEC) Program through the NASA Advanced Supercomputing (NAS) Division at Ames Research Center for the production of the SPOC data products.

This work makes use of observations from the LCOGT network. Part of the LCOGT telescope time was granted by NOIRLab through the Mid-Scale Innovations Program (MSIP). MSIP is funded by NSF.

This work is based in part on observations obtained at the Southern Astrophysical Research (SOAR) telescope, which is a joint project of the Minist\'{e}rio da Ci\^{e}ncia, Tecnologia e Inova\c{c}\~{o}es (MCTI/LNA) do Brasil, the US National Science Foundation’s NOIRLab, the University of North Carolina at Chapel Hill (UNC), and Michigan State University (MSU).

This study is based on observations collected at the European Southern Observatory under ESO programme 0103.C-0719.

The authors acknowledge the Texas Advanced Computing Center (TACC) at The University of Texas at Austin for providing HPC resources that have contributed to the research results reported within this paper\footnote{\url{http://www.tacc.utexas.edu}}.

This research made use of the open source Python package {\tt exoctk}, the Exoplanet Characterization Toolkit \citep{Bourqueetal2021}.

The national facility capability for SkyMapper has been funded through ARC LIEF grant LE130100104 from the Australian Research Council, awarded to the University of Sydney, the Australian National University, Swinburne University of Technology, the University of Queensland, the University of Western Australia, the University of Melbourne, Curtin University of Technology, Monash University, and the Australian Astronomical Observatory. SkyMapper is owned and operated by The Australian National University's Research School of Astronomy and Astrophysics. The survey data were processed and provided by the SkyMapper Team at ANU.

Figures in this manuscript were created using color-impaired-friendly schemes from ColorBrewer 2.0\footnote{\url{https://colorbrewer2.org/}}

We would like to acknowledge the Alabama-Coushatta, Caddo, Carrizo/Comecrudo, Coahuiltecan, Comanche, Kickapoo, Lipan Apache, Tonkawa and Ysleta Del Sur Pueblo, and all of the American Indian and Indigenous Peoples and communities who have been or have become a part of these lands and territories in Texas.

\facilities{SALT--HRS, LCO--Sinistros, HARPS, Texas Advanced Computing Center (TACC)}

\software{{\tt AstroImageJ} \citep{Collinsetal2017},
{\tt astropy} \citep{astropy1,astropy2}, 
{\tt celerite} \citep{Foreman-Mackeyetal2017}, 
{\tt comove} \citep{Tofflemireetal2021},
{\tt emcee} \citep{Foreman-Mackeyetal2013}, 
{\tt misttborn} \citep{Mannetal2016a,Johnsonetal2018},
{\tt saphires} \citep{Tofflemireetal2019},
{\tt scipy} \citep{scipy,scipy2}.
}


\end{document}